\documentclass[twocolumn,superscriptaddress,nobalancelastpage,10pt,pra,letter]{revtex4}
\usepackage{amsmath}
\usepackage{amsthm}
\usepackage{amssymb}
\usepackage{amsfonts}
\usepackage{graphicx}
\usepackage{wasysym}
\usepackage{mathrsfs}
\usepackage{yfonts}
\usepackage{bbold}
\usepackage{verbatim}
\usepackage{subfigure}
\usepackage{color}
\usepackage{setspace}

\makeatletter
\let\save@mathaccent\mathaccent
\newcommand*\if@single[3]{%
	\setbox0\hbox{${\mathaccent"0362{#1}}^H$}%
	\setbox2\hbox{${\mathaccent"0362{\kern0pt#1}}^H$}%
	\ifdim\ht0=\ht2 #3\else #2\fi
}
\newcommand*\rel@kern[1]{\kern#1\dimexpr\macc@kerna}
\newcommand*\widebar[1]{\@ifnextchar^{{\wide@bar{#1}{0}}}{\wide@bar{#1}{1}}}
\newcommand*\wide@bar[2]{\if@single{#1}{\wide@bar@{#1}{#2}{1}}{\wide@bar@{#1}{#2}{2}}}
\newcommand*\wide@bar@[3]{%
	\begingroup
	\def\mathaccent##1##2{%
		\let\mathaccent\save@mathaccent
		\if#32 \let\macc@nucleus\first@char \fi
		\setbox\z@\hbox{$\macc@style{\macc@nucleus}_{}$}%
		\setbox\tw@\hbox{$\macc@style{\macc@nucleus}{}_{}$}%
		\dimen@\wd\tw@
		\advance\dimen@-\wd\z@
		\divide\dimen@ 3
		\@tempdima\wd\tw@
		\advance\@tempdima-\scriptspace
		\divide\@tempdima 10
		\advance\dimen@-\@tempdima
		\ifdim\dimen@>\z@ \dimen@0pt\fi
		\rel@kern{0.6}\kern-\dimen@
		\if#31
		\overline{\rel@kern{-0.6}\kern\dimen@\macc@nucleus\rel@kern{0.4}\kern\dimen@}%
		\advance\dimen@0.4\dimexpr\macc@kerna
		\let\final@kern#2%
		\ifdim\dimen@<\z@ \let\final@kern1\fi
		\if\final@kern1 \kern-\dimen@\fi
		\else
		\overline{\rel@kern{-0.6}\kern\dimen@#1}%
		\fi
	}%
	\macc@depth\@ne
	\let\math@bgroup\@empty \let\math@egroup\macc@set@skewchar
	\mathsurround\z@ \frozen@everymath{\mathgroup\macc@group\relax}%
	\macc@set@skewchar\relax
	\let\mathaccentV\macc@nested@a
	\if#31
	\macc@nested@a\relax111{#1}%
	\else
	\def\gobble@till@marker##1\endmarker{}%
	\futurelet\first@char\gobble@till@marker#1\endmarker
	\ifcat\noexpand\first@char A\else
	\def\first@char{}%
	\fi
	\macc@nested@a\relax111{\first@char}%
	\fi
	\endgroup
}
\makeatother

\newcommand{\ket}[1]{| #1 \rangle}
\newcommand{\bra}[1]{\langle #1 |}

\newcommand{\dm}{\widehat{\rho}}

\newcommand{\avgr}[2]{\text{tr}\left\{ \widehat{\rho}_{#1} #2 \right\}}

\newcommand{\opa}{\widehat{a}}

\newcommand{\opEps}{\widehat{E}^{(+)}}
\newcommand{\opEns}{\widehat{E}^{(-)}}

\newcommand{\I}{\mathscr{I}}

\renewcommand{\L}{\mathscr{L}}

\makeatletter
\newcommand*{\rom}[1]{\expandafter\@slowromancap\romannumeral #1@}
\makeatother

\makeatletter
\newcommand{\roml}[1]{\lowercase\expandafter{\romannumeral #1\relax}}
\makeatother

\DeclareMathAlphabet{\mathpzc}{OT1}{pzc}{m}{it}

\begin{document}
	
	\title{Single-qubit measurement of two-qubit entanglement in generalized Werner states}
	
	\author{Salini Rajeev}
	\affiliation{Department of Physics, 145 Physical Sciences Bldg., Oklahoma State University, Stillwater, OK 74078, USA.}
	
	\author{Mayukh Lahiri}
	\email{mlahiri@okstate.edu} \affiliation{Department of Physics, 145 Physical Sciences Bldg., Oklahoma State University, Stillwater, OK 74078, USA.}

	\begin{abstract}
		Conventional methods of measuring entanglement in a two-qubit photonic mixed state require the detection of both qubits. We generalize a recently introduced method which does not require the detection of both qubits, by extending it to cover a wider class of entangled states. Specifically, we present a detailed theory that shows how to measure entanglement in a family of two-qubit mixed states \textemdash~obtained by generalizing Werner states \textemdash~without detecting one of the qubits. Our method is interferometric and does not require any coincidence measurement or postselection. We also perform a quantitative analysis of anticipated experimental imperfections to show that the method is experimentally implementable and resistant to experimental losses.
	\end{abstract}
	
	\maketitle
	\section{Introduction}\label{sec:intro}
Traditional methods of characterizing entanglement in two-photon mixed states (e.g., the violation of Bell's inequalities \cite{freedman1972experimental,aspect1982experimental,ou1992realization,giustina-2015-significant}, quantum state tomography \cite{James2001}, etc.) require detection of both photons (see, for example, \cite{guhne2009entanglement,friis2019entanglement} and references therein). These methods, therefore, involve coincidence measurement or postselection. Methods that do not require detection of both photons rely on the assumption that the quantum state is pure (see, for example, \cite{walborn2006,sahoo2020quantum,bhattacharjee2022measurement,Pires2009,Just2013,sharapova2015schmidt}).
\par
Recently, it has been demonstrated theoretically \cite{pol-ent-theory} and experimentally \cite{pol-ent-exp} that it is possible to measure entanglement of a special class of two-photon mixed states \textemdash ~obtained by generalizing Bell states \textemdash ~without detecting one of the photons and without employing coincidence measurement or postselection. Density matrices representing such states have two generally non-vanishing coherence terms (off-diagonal elements). The states are entangled if and only if these coherence terms are nonzero. Furthermore, these two coherence terms are complex conjugate of each other. Therefore, entanglement of the states considered in Refs.~\cite{pol-ent-theory,pol-ent-exp} is fully characterized by one coherence term of the corresponding density matrix. However, entanglement of most two-qubit states are not dependent on their density matrix elements in such a trivial manner, e.g., the Werner state that can be entangled without violating Bell's inequalities \cite{Werner1989_PRA}. Whether the entanglement of any two-qubit photonic mixed state can be verified by detecting one qubit remains an open question of fundamental importance.
\par
Here, we take an important step toward answering this question by extending the method to two-qubit mixed states that can be obtained by generalizing Werner states. We find that albeit the same principle introduced in Refs.\,\cite{pol-ent-theory,pol-ent-exp} applies, the measurement procedure requires considerable modifications. Our results also suggest that the method to account for experimental losses would require significant adaptation (Supplementary Material). 
\par
The article is organized as follows: In Sec.\,\ref{sec:q-state}, we discuss the class of quantum state we address and their entanglement. In Sec.\,\ref{sec:meas-scheme}, we provide an outline of the entanglement measurement scheme. In Sec.\,\ref{sec:theory} we present a detailed theoretical analysis and our main results. We also illustrate the results by numerical examples. In Sec.\,\ref{sec:comp} we compare our results with existing ones. In Sec.\,\ref{sec:exp-losses}, we discuss the effect of anticipated experimental imperfections. Finally, we summarize and conclude in Sec.\,\ref{sec:summary}.

\section{Two-Qubit Generalized Werner State and Its Entanglement}\label{sec:q-state}
\begin{figure*}
	\centering
	\includegraphics[width=0.98\linewidth]{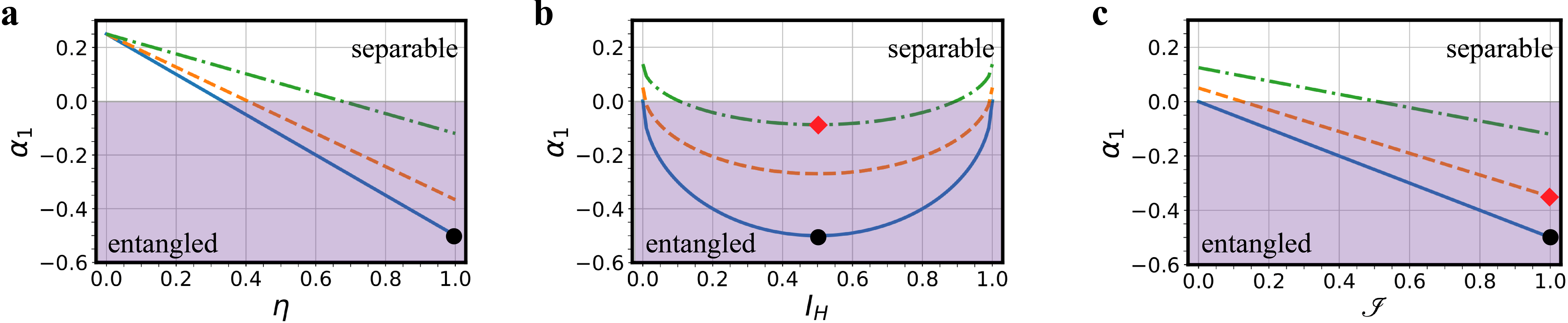}
	\qquad
	\caption{Illustration of the PPT criterion applied to generalized Werner states. The state is entangled when the eigenvalue $\alpha_1<0$ (shaded regions). The black filled circle and the red diamond represent a Bell state ($\ket{\Phi^+}$ or $\ket{\Phi^-}$) and a Werner state, respectively. \textbf{a}, $\alpha_1$ is plotted against $\eta$ for $\I=1$ and $I_H=0.5$ (solid line), for $\I=0.8$ and $I_H=0.3$ (dashed line), and for $\I=0.3$ and $I_H=0.8$ (dash-dotted line). \textbf{b}, $\alpha_1$ is plotted against $I_H$ for $\I=1$ and $\eta=1$ (solid line), for $\I=0.8$ and $\eta=0.8$ (dashed line), and for $\I=1$ and $\eta=0.45$ (dash-dotted line). {\bf c}, $\alpha_1$ is plotted against $\I$ for $I_H=0.5$ and $\eta=1$ (solid line), $I_H=0.5$ and $\eta=0.8$ (dashed line), and $I_H=0.6$ and $\eta=0.5$ (dash-dotted line).} \label{fig:PPT-fig}
\end{figure*}
We work with a two-photon polarization state that is a common test bed for two-qubit systems. Throughout this paper, we call the two photons forming a pair signal ($S$) and idler ($I$). We use $H,V,D,A,R$, and $L$ to represent horizontal, vertical, diagonal, antidiagonal, right-circular, and left-circular polarization, respectively. The ket, $\ket{\mu_I,\nu_S}$, represents a photon pair where the idler photon has polarization $\mu$ and the signal photon has polarization $\nu$.
\par
The quantum state considered here is obtained by generalizing the Werner state. The density matrix takes the following form in the computational basis $\{ \ket{H_IH_S}, \ket{H_IV_S}, \ket{V_IH_S}, \ket{V_IV_S} \}$: 
\begin{align}\label{q-state}
\dm=\begin{pmatrix}
\eta I_H+\frac{1-\eta}{4} & 0 & 0 & \eta \mathscr{I}\sqrt{I_HI_V}e^{-i\phi} \\
0 & \frac{1-\eta}{4} & 0 & 0\\
0 & 0 & \frac{1-\eta}{4} & 0\\
\eta \mathscr{I}\sqrt{I_HI_V}e^{i\phi} & 0 & 0 & \eta I_V+\frac{1-\eta}{4}
\end{pmatrix},
\end{align}
where $0\leq I_H \leq 1$, $I_V=1-I_H$, $0\leq \eta \leq 1$, $0\leq \I \leq 1$, and $\phi$ represents a phase. It can be immediately checked that this state becomes a Werner state when $I_H=I_V=1/2$ and $\I=1$.  For $I_H=I_V=1/2$, $\I=1$ and $\eta=1$, it reduces to Bell states $\ket{\Phi^+}=(\ket{H_IH_S}+\ket{V_IV_S})/\sqrt{2}$ and $\ket{\Phi^-}=(\ket{H_IH_S}-\ket{V_IV_S})/\sqrt{2}$ when $\phi=0$ and $\phi=\pi$, respectively. Therefore, the state given by Eq.~(\ref{q-state}) can also be obtained by the convex combination of a fully mixed state with the generalization of Bell states considered in Refs.\,\cite{pol-ent-theory,pol-ent-exp}. 
\par
The entanglement of this state can be verified by testing the \emph{positive partial transpose} (PPT) criterion \cite{Peres-sep-cond-PRL}. A partial transposition of the  density matrix ($\widehat{\rho}$) is a transposition taken with respect to only one of the photons. The density matrix has a positive partial transpose if and only if its partial transposition does not have any negative eigenvalues. Since we have a $2\times 2$ system, according to the PPT criterion the state is entangled if and only if it does not have a positive partial transpose \cite{Horodeckis-separability-PLA}. 
\par
By determining the eigenvalues of a partial transpose of the density matrix given by Eq.~(\ref{q-state}), we find that three of them are always positive (see Appendix A). The only eigenvalue that can be negative is given by
\begin{equation}\label{PPT-eval}
\alpha_{1}=\frac{1-\eta-4\eta\mathscr{I}\sqrt{I_HI_V}}{4}.
\end{equation} 
Figure \ref{fig:PPT-fig} illustrates entanglement of generalized Werner states ($\dm$) characterized by the PPT criterion for various choices of state parameters $\eta$, $I_H$, and $\I$.
\par
The amount of entanglement present in the quantum state can be quantified by the concurrence \cite{wootters1998entanglement}. We find that the concurrence of the state given by Eq.~(\ref{q-state}) is (see Appendix B) 
\begin{align}\label{conc-form}
C(\widehat{\rho})=\text{max}\left\{\frac{\eta-1+4\eta\mathscr{I}\sqrt{I_HI_V}}{2},0\right\}.
\end{align}
We show below how to determine the concurrence without detecting one of the photons.

\section{Outline of the Entanglement Measurement Scheme}\label{sec:meas-scheme}
The principle of our method is based on a unique quantum interference phenomenon that was first demonstrated by Zou, Wang, and Mandel \cite{zou1991induced,wang1991induced} and is sometimes called the interference by path identity \cite{Lahiri2022PI_RMP}. The method employs a nonlinear interferometer \cite{chekhova2016nonlinear} that contains two identical twin-photon sources. Each source produces the same quantum state [Eq.~(\ref{q-state})].
\par
\begin{figure}[b]  \centering
	\includegraphics[width=0.98\linewidth]{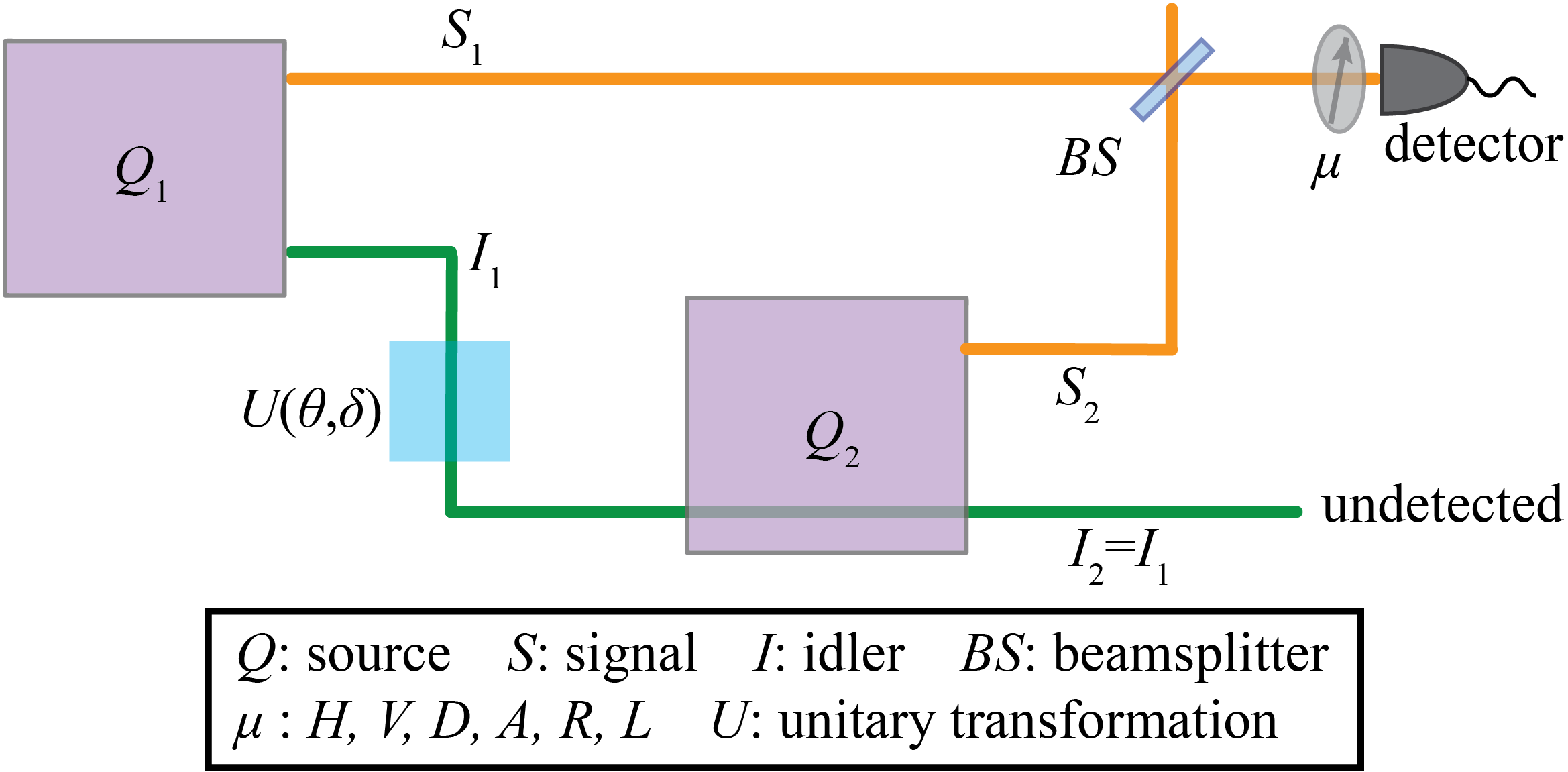}
	\qquad
	\caption{Entanglement measurement scheme. Two sources ($Q_1$ and $Q_2$) can individually generate the two-qubit photonic state $\dm$ [Eq.~(\ref{q-state})]. $Q_1$ emits a photon pair, signal and idler, into propagation modes $S_1$ and $I_1$. Likewise, $Q_2$ emits signal and idler photons into modes $S_2$ and $I_2$. Modes $S_1$ and $S_2$ are combined by a beamsplitter ($BS$) and the signal photon emerging from $BS$ is sent to a detector after projecting it onto a chosen polarization state ($\mu=H$, $V$, $D$, $A$, $R$, or $L$). Modes $I_1$ and $I_2$ are made identical by aligning the corresponding beams without the use of a beamsplitter. This alignment makes it impossible to know from which source a signal photon originated and consequently, a single-photon interference pattern appears at the detector. The idler photon is never detected and no postselection is considered to obtain the interference pattern. A unitary transformation, $U(\theta,\delta)$, is performed on the idler photon between the two sources. The information about the entanglement is extracted from the interference patterns with the knowledge of this transformation.} \label{fig:scheme}
\end{figure}

A conceptual arrangement of the entanglement measurement scheme is illustrated in Fig.~\ref{fig:scheme}. The two photon-pair sources are denoted by $Q_1$ and $Q_2$. A photon pair is in superposition of being created at the two sources \textemdash ~for sources made of nonlinear crystals, such a situation can be achieved by weakly pumping them with mutually coherent laser beams. The sources do not produce the states simultaneously, i.e., the probability of having more than one photon pair in the system between an emission and a detection is negligible. That is, in this situation the effect of stimulated (induced) emission is negligible \cite{zou1991induced,wang1991induced,wiseman2000induced,liu2009investigation,lahiri2019nonclassicality}. 
\par
Signal beams ($S_1$ and $S_2$) from the two sources are superposed by a balanced beamsplitter and the single-photon counting rate (intensity) is measured at one of the outputs of the beamsplitter. The idler beam from $Q_1$ is sent through $Q_2$ and is aligned with the idler beam generated by $Q_2$. (Such an alignment was originally suggested by Z. Y. Ou \cite{zou1991induced}.) This alignment makes it impossible to know from which source the signal photon originated. Consequently, a single-photon interference pattern appears at a detector placed at an output of the beamsplitter. Details of experimental conditions to obtain the interference pattern have been discussed in numerous publications (see, for example, \cite{wang1991induced,lemos2014quantum,lemos2022quantum}).
\par
We apply a unitary transformation on the field representing the idler photon between the two sources. Such transformations can readily be implemented in a laboratory using quarter- and half-wave plates. We choose a unitary transformation that is characterized by two controllable parameters ($\delta$ and $\theta$) and has the following form in the $\{ \ket{H}, \ket{V}\}$ basis:
\begin{align}\label{idler-transform}
U(\theta,\delta)=\begin{pmatrix}
e^{-i\delta}\cos 2\theta & e^{-i\delta}\sin 2\theta \\
e^{i\delta}\sin 2\theta & -e^{i\delta}\cos 2\theta 
\end{pmatrix},
\end{align} 
where $0\leq \theta \leq \pi$ and $0\leq \delta/2 \leq \pi$ can be understood as two half-wave plate angles. 
\par
The fact that we can fully control the choice of $\theta$ and $\delta$ plays crucial role in our measurement scheme. We show below that the effect of this unitary transformation appears in the interference patterns generated by the signal photon after it is projected onto appropriate polarization states. We also show how the entanglement of $\dm$ can be fully characterized from these single-photon interference patterns with the knowledge of the unitary transformation. For simplicity, we initially assume that there is no experimental loss. In the supplementary material, we provide a detailed description of how to treat dominant experimental losses and imperfections. We discuss the effects of experimental imperfections in Sec.\,\ref{sec:exp-losses}. 
\par
We emphasize that \emph{the signal photon never interacts with the device performing the unitary transformation and the idler photon is never detected}. These are two unique features of our entanglement measurement scheme in addition to the fact that \emph{no postselection or coincidence measurement is required}.

\section{Theoretical Analysis}\label{sec:theory}
\subsection{Determining the quantum state}\label{subsec:q-state}
We use the standard bra-ket notation for the convenience of analysis. The generalized Werner state [Eq.~(\ref{q-state})] in the bra-ket notation takes the form
\begin{align}\label{q-state-2}
\dm=\sum_{\mu,\nu}^{H,V} \sum_{\mu',\nu'}^{H,V} \bra{\mu_I\nu_S}\dm\ket{\mu_I'\nu_S'} \ket{\mu_I\nu_S} \bra {\mu_I'\nu_S'},
\end{align}
where $\bra{\mu_I\nu_S}\dm\ket{\mu_I'\nu_S'}$ represents a matrix element; for example, $\bra{H_I H_S}\dm\ket{H_I H_S}=\eta I_H + (1-\eta)/4$.
\par
Equation (\ref{q-state-2}) [or equivalently, Eq.~(\ref{q-state})] is the state generated by an individual source. While determining the quantum state generated by the two sources jointly, one needs to use the fact that the probability of emission of more than one photon pair is negligible, that is, the total occupation number of photons in the state is always two. 
\par
We first consider the scenario in which the idler beams are not aligned. In this case, a signal photon is in a superposition of being in modes $S_1$ and $S_2$ that emerges from two sources. Likewise, an idler photon is in a superposition of being in modes $I_1$ and $I_2$. Consequently, the density matrix representing the quantum state produced jointly by the two sources becomes an $8\times 8$ matrix. We determine this matrix following the approach introduced in Ref.\,\cite{pol-ent-theory} and represent it by bra-ket notation [Appendix C, Eq.~(\ref{q-st-two-source})].
\par
We now analytically represent the alignment of idler beams. Modes $I_1$ and $I_2$ becomes identical due to this alignment. The alignment, together with the unitary transformation [Eq.~(\ref{idler-transform})] applied on the idler photon, results in the following transformations of kets:
\begin{subequations}\label{align-cond}
	\begin{align}
	&\ket{H_{I_2}}=e^{-i\phi_I} \left(e^{i\delta}\cos 2\theta \ket{H_{I_1}} + e^{i\delta}\sin 2\theta \ket{V_{I_1}} \right), \label{align-cond-a}\\
	& \ket{V_{I_2}}=e^{-i\phi_I} \left(e^{-i\delta}\sin 2\theta  \ket{H_{I_1}} - e^{-i\delta}\cos 2\theta  \ket{V_{I_1}} \right) \label{align-cond-b},
	\end{align} 
\end{subequations}
where $\phi_I$ is the phase acquired by propagation from $Q_1$ to $Q_2$. 
\par
Finally, we obtain the density matrix, $\dm^{(f)}$, representing the photon pair in our system (before arriving at $BS$) by substituting from Eqs.~(\ref{align-cond-a}) and (\ref{align-cond-b}) into the $8\times 8$ density matrix [Appendix C, Eq.~(\ref{q-st-two-source})] mentioned above. An expression for $\dm^{(f)}$ is given by Eq.~(\ref{dm-final-exp}) in Appendix C. The probability of detecting a signal photon at one of the outputs of $BS$ (Fig.\,\ref{fig:scheme}) can be determined using this density matrix or from the reduced density matrix ($\dm_S$) obtained by taking partial trace of $\dm^{(f)}$ over the subspace of the idler photon. Here, we take the later approach. The reduced density matrix representing a signal photon before arriving at $BS$ is given by 
\begin{widetext}
	\begin{align}\label{signal-dm}
	\dm_S &=K(\eta,I_H) \left( |b_1|^2 \ket{H_{S_1}}\bra{H_{S_1}} +|b_2|^2 \ket{H_{S_2}}\bra{H_{S_2}} \right) +K(\eta,I_V) \left( |b_1|^2 \ket{V_{S_1}}\bra{V_{S_1}} +|b_2|^2 \ket{V_{S_2}}\bra{V_{S_2}} \right)  \nonumber \\ & +\Big\{ b_1 b_2^{\ast} e^{i\phi_I} \big[ \left\{ \L(\eta,I_H,\delta)\ket{H_{S_1}}\bra{H_{S_2}} +\L'(\eta,I_V,\delta)\ket{V_{S_1}}\bra{V_{S_2}} \right\} \cos 2\theta \nonumber \\ & \qquad \qquad \quad + \eta \I \sqrt{I_HI_V} \left\{ \Phi(\delta) \ket{H_{S_1}}\bra{V_{S_2}} +\Phi'(\delta) \ket{V_{S_1}}\bra{H_{S_2}}  \right\} \sin 2\theta \big] +\text{H.c.} \Big\},
	\end{align}
\end{widetext}
where $|b_j|^2$ is the probability of emission from source $Q_j$, H.c.~represents the Hermitian conjugation, $K(\eta,I)=\eta I+(1-\eta)/2$, $\Phi(\delta)=\exp[i(\phi^{V_2,V_2}_{H_1,H_1}+\delta)]$, $\Phi'(\delta)=\exp[i(\phi^{H_2,H_2}_{V_1,V_1}-\delta)]$, and 
\begin{subequations}\label{signal-dm-exp}
	\begin{align}
	\L(\eta,I_H,\delta)=& M(\eta,I_H) \exp \left[i \left(\phi^{H_2,H_2}_{H_1,H_1}-\delta \right) \right] \nonumber \\ &-N(\eta)\exp \left[i \left(\phi^{V_2,H_2}_{V_1,H_1}+\delta \right) \right], \\ \L'(\eta,I_V,\delta)=& N(\eta)\exp \left[ i \left(\phi^{H_2,V_2}_{H_1,V_1}-\delta \right) \right]  \nonumber \\ & -M(\eta,I_V) \exp \left[ i \left(\phi^{V_2,V_2}_{V_1,V_1}+\delta \right) \right],
	\end{align}
\end{subequations}	
with $M(\eta,I)=(4\eta I+1-\eta)/4$ and $N(\eta)=(1-\eta)/4$. We observe that all state parameters ($\eta$, $I_H$, and $\I$) and the unitary transformation parameters ($\theta$ and $\delta$) appear in the density matrix representing a signal photon.

\subsection{Information of entanglement in Interference Patterns}\label{subsec:ent-vis}
As mentioned in Sec.~\ref{sec:meas-scheme}, signal beams $S_1$ and $S_2$ are superposed by a balanced beamsplitter (BS), one of the outputs of BS is projected onto a particular polarization state, and then sent to a detector. Therefore, the positive-frequency part of the quantized electric field at the detector is given by	
\begin{align}\label{field-at-detector}
\opEps_{\mu_S}=\frac{1}{\sqrt{2}}\big(\opa_{\mu_{S1}}+ie^{i\phi_S}\opa_{\mu_{S2}}\big),
\end{align}
where $\mu=H,V,D,A,R,L$ and $\opa_{\mu_{Sj}}$ is the annihilation operator corresponding to signal photon with polarization $\mu$ emitted by source $Q_j$. The probability of detecting a signal photon with polarization $\mu$ is obtained by the standard formula \cite{glauber1963quantum}
\begin{align}\label{ph-count-rt}
P_{\mu}=\avgr{S}{\opEns_{\mu_S}\opEps_{\mu_S}},
\end{align}
where $\opEns_{\mu_S}=\{\opEps_{\mu_S}\}^{\dag}$ and $\dm_S$ is given by Eq.~(\ref{signal-dm}). The single-photon counting rate (intensity) at the detector is linearly proportional to the probability $P_{\mu}$. We show below that these photon counting rates represent interference patterns. The state parameters ($\eta, I_H, \I$) that  characterize the entanglement appear in the expressions for these interference patterns.
\par
We first determine the photon counting rate at the detector when the signal photon is projected onto the horizontally polarized (H) state. It follows from Eqs.~(\ref{signal-dm}) -- (\ref{ph-count-rt}) that (Appendix D)
\begin{align}\label{ph-counting-rates-h}
P_{H}\big|_{\theta=0}& = \frac{1}{2}\big[P_1+P_2+2 |b_1| |b_2| \sin \left(\phi_{\text{in}}+\phi_0 \right) \nonumber \\ &\times \{ P_1^2+P_2^2 -2P_1P_2 \cos(\chi + 2\delta)\}^{\frac{1}{2}}\big], 
\end{align}
where $\phi_{\text{in}}$=$\arg\{b_1\}-\arg\{b_2\}+\phi_I-\phi_S$, $P_1=\eta I_{H}+(1-\eta)/4$, $P_2=(1-\eta)/4$, $\chi=\phi^{V_2H_2}_{V_1H_1}-\phi^{H_2H_2}_{H_1H_1}$ and $\phi_0=\phi_{H_1H_{1}}^{H_2H_2}-\delta+\epsilon_1$; the explicit form of $\epsilon_1$ is not required for our purpose. It is evident from Eq.~(\ref{ph-counting-rates-h}) that the value of $P_{H}\big|_{\theta=0}$ changes sinusoidally as $\phi_{\text{in}}$ is varied, i.e., Eq.~(\ref{ph-counting-rates-h}) represents an interference pattern. Here, we have chosen $\theta=0$ to maximize the contribution from the interference term (general expressions are given in Appendix D). 
\par
Equation (\ref{ph-counting-rates-h}) shows that the visibility depends on the parameter $\delta$ that one can fully control in an experiment. We set $\delta=\delta_H$ such that the visibility attains its minimum non-zero value, i.e, $\cos (\chi+2\delta_H)=1$. The expression for $P_H$ then becomes
\begin{align}\label{ph-countng-rate-h-final}
P_{H}\big|_{\theta=0}^{\delta=\delta_H}= \frac{1-\eta}{4}+\frac{\eta I_{H}}{2}+|b_1| |b_2| \eta I_{H} \sin (\phi_\text{in}+\phi_0),
\end{align}
and consequently, the visibility is given by  \footnote{The visibility is determined by the standard formula $\mathcal{V}_{\mu}=\left(P^{\text{max}}_{\mu}
	-P^{\text{min}}_{\mu}\right) / \left(P^{\text{max}}_{\mu}
	+P^{\text{min}}_{\mu}\right)$.}
\begin{align}\label{vis-def-h}
\mathcal{V}_H\big|_{\theta=0}^{\delta=\delta_H}=\frac{4|b_1| |b_2|\eta I_H}{2\eta I_H+1-\eta}.
\end{align}
Likewise, we find that the visibility of the single-photon interference pattern for $V$ polarization is
\begin{align}\label{vis-def-v}
\mathcal{V}_V\big|_{\theta=0}^{\delta=\delta_V}=\frac{4|b_1| |b_2|\eta(1-I_H)}{1+\eta-2\eta I_H},
\end{align}
where $\delta_V$ plays the same role as $\delta_H$ in Eq.~(\ref{vis-def-h}).
\begin{table*}[t]		
	\setlength{\tabcolsep}{10pt} 
	\renewcommand{\arraystretch}{1} 
	\begin{tabular}{c  c  c  c  c  c  c  c} 
		\hline\hline 
		State & ($\eta,\I,I_H$) & $\mathcal{V}_R|_{\theta=\frac{\pi}{4}}$  &  $\mathcal{V}_D|_{\theta=\frac{\pi}{4}}$  &
		$\mathcal{V}_V\big|_{\theta=0}^{\delta=\delta_V}$ &
		$\mathcal{V}_H\big|_{\theta=0}^{\delta=\delta_H}$ &
		PPT Criterion & Concurrence \\[2 pt] 
		\hline 
		$\widehat{\rho}_1$ & (0.0, --, --) & 0.00 & 0.00 & 0.00 & 0.00 & Separable & 0.00 \\
		$\widehat{\rho}_2$ & (0.2, 1.0, 0.5) & 0.08 & 0.18 & 0.20 & 0.20 & Separable & 0.00 \\
		$\widehat{\rho}_3$ & (0.6, 0.8, 0.3) & 0.17 & 0.41 & 0.67 & 0.47 & Entangled& 0.24 \\ 
		$\widehat{\rho}_4$ & (0.7, 1.0, 0.5) & 0.27 & 0.64 & 0.70 & 0.70 & Entangled & 0.55  \\
		$\widehat{\rho}_5$ & (1.0, 1.0, 0.5) & 0.38 & 0.92 & 1.00 & 1.00 & Entangled & 1.00 \\ [1ex] 
		\hline\hline
	\end{tabular}
	\caption{Numerical results illustrating entanglement verification for five quantum states through single-photon interference. }\label{tab:1}
\end{table*}
\par
We now consider the remaining cases where the signal photon is projected onto diagonal (D), antidiagonal (A), right-circular (R), and left-circular (L) polarizations. General expressions for single-photon counting rates for all these cases are given in Appendix D. Using those expressions [Eqs.~(\ref{ph-coutng-rate-DARL:a})--(\ref{ph-coutng-rate-DARL:d})], we obtain the corresponding formulas for visibility as
\begin{align}
&\mathcal{V}_D\big|_{\theta=\frac{\pi}{4}} =\mathcal{V}_A\big|_{\theta=\frac{\pi}{4}} \nonumber \\ &= 2|b_1| |b_2| \eta \mathscr{I}\sqrt{I_HI_V}\sqrt{2+2\cos (\chi'-2\delta)}, \label{vis-dr:a} 
\end{align}
and
\begin{align}
&\mathcal{V}_R\big|_{\theta=\frac{\pi}{4}} =\mathcal{V}_L\big|_{\theta=\frac{\pi}{4}} \nonumber \\ & =2 |b_1| |b_2| \eta \mathscr{I}\sqrt{I_HI_V}\sqrt{2-2\cos (\chi'-2\delta)}, \label{vis-dr:b}
\end{align}
where $\chi'$ is given in Appendix D (its explicit form is not required for our purpose). 
\par 
Equations (\ref{vis-def-h})--(\ref{vis-dr:b}) show that all parameters that characterize the entanglement appear in the expressions for visibility for $H$, $V$, $D$, $A$, $R$, and $L$ polarizations. That is, the information about the entanglement is contained in the single-photon interference patterns obtained by detecting one of the qubits only. 

\subsection{Entanglement Verification and Measurement}\label{subsec:PPT}
We now show how to test the PPT criterion and to measure the concurrence from the single-photo interference patterns. It follows from Eqs.~(\ref{PPT-eval}) and (\ref{conc-form}) that it is enough to represent $\eta$ and $\eta \mathscr{I}\sqrt{I_H I_V}$ in terms of experimentally measurable quantities. We show below that these two quantities can be directly obtained from the expressions for visibilities given in Sec.\,\ref{subsec:ent-vis}. That is, we do not need to determine all state parameters ($\eta$, $I_H$, and $\I$) individually. 
\par
We first express $|b_1| |b_2|$ in terms of experimentally measurable quantities. We recall that $|b_1|^2$ and $|b_2|^2$ are probabilities of emission from sources $Q_1$ and $Q_2$, respectively. Suppose that $P_{\mu}^{(1)}$ is the probability of detecting a signal photon with polarization $\mu$ ($\mu=H,V$) when signal beam, $S_2$, emerging from $Q_2$ is blocked. Likewise, $P_{\mu}^{(2)}$ is the detection-probability when signal beam $S_1$ is blocked. We thus have  
\begin{align}\label{b1b2}
|b_1| |b_2| =\frac{\sqrt{P_{\mu}^{(1)}P_{\mu}^{(2)}}}{P_{\mu}^{(1)}+P_{\mu}^{(2)}}.
\end{align}
\par
We now apply Eqs.~(\ref{vis-def-h}) and (\ref{vis-def-v}) to determine $\eta$ and find that
\begin{align}\label{eta}
\eta=\frac{ \mathcal{V}_V\big|_{\theta=0}^{\delta=\delta_V}+\mathcal{V}_H\big|_{\theta=0}^{\delta=\delta_H} -\frac{1}{|b_1| |b_2|}\mathcal{V}_H\big|_{\theta=0}^{\delta=\delta_H}\mathcal{V}_V\big|_{\theta=0}^{\delta=\delta_V}}{4 |b_1| |b_2| -\mathcal{V}_V\big|_{\theta=0}^{\delta=\delta_V}-\mathcal{V}_H\big|_{\theta=0}^{\delta=\delta_H}}.
\end{align}
It can be checked that when $|b_1|= |b_2|=1/\sqrt{2}$, the right-hand side of Eq.~(\ref{eta}) becomes equal to 1 in the limit $\mathcal{V}_H\big|_{\theta=0}^{\delta=\delta_H}\to 1-$ and $\mathcal{V}_V\big|_{\theta=0}^{\delta=\delta_V}\to 1-$. In this limit, the generalized Werner state [Eq.~(\ref{q-state})] reduces to the generalization of Bell states considered in Refs.\,\cite{pol-ent-theory,pol-ent-exp}, for which $\eta=1$. Using Eqs.~(\ref{b1b2}) and (\ref{eta}) one can immediately express $\eta$ in terms of experimentally measurable quantities.
\par
To determine the quantity $\eta \mathscr{I}\sqrt{I_H I_V}$, we eliminate the cosine terms from Eqs.~(\ref{vis-dr:a}) and (\ref{vis-dr:b}) by squaring and adding them. It then immediately follows that 
\begin{align}\label{vdvr-combn}
\eta \mathscr{I}\sqrt{I_H I_V}= \frac{1}{4|b_1| |b_2|} \sqrt{(\mathcal{V}_D|_{\theta=\frac{\pi}{4}})^2+(\mathcal{V}_R|_{\theta=\frac{\pi}{4}})^2}.
\end{align}
Using Eqs.~(\ref{b1b2}) and (\ref{vdvr-combn}), we can readily express $\eta \mathscr{I}\sqrt{I_H I_V}$ in terms of photon counting rates and visibilities of single-photon interference patterns. 
\par
To test the PPT criterion, we express the eigenvalue, $\alpha_1$ [Eq.~(\ref{PPT-eval})], in terms of experimentally measurable quantities. Using Eq.~(\ref{PPT-eval}) and Eqs.~(\ref{b1b2})--(\ref{vdvr-combn}), we find that
\begin{align}\label{eigenval-vis}
\alpha_1= \frac{1-\mathcal{V}_{HV}-4\mathcal{V}_{DR} }{4},
\end{align}
where we have denoted the right-hand sides of Eqs.~(\ref{eta}) and (\ref{vdvr-combn}) by $\mathcal{V}_{HV}$ and $\mathcal{V}_{DR}$, respectively. According to the PPT criterion, the state $\dm$ [Eq.~(\ref{q-state})] is entangled if and only if $\alpha_1<0$, that is, the state is entangled if and only if
\begin{align}\label{PPT-vis}
\mathcal{V}_{HV}+4\mathcal{V}_{DR} >1.
\end{align}
\par
We illustrate the PPT criterion by choosing five quantum states ($\dm_1,\dots,\dm_5$) in Table\,\ref{tab:1}. The criterion is tested using Eq.~(\ref{PPT-vis}). For simplicity, we assume that $|b_1|= |b_2|=1/\sqrt{2}$. To compute visibilities for $D$ and $R$ polarization, we chose $\chi'-2\delta=\pi/4$. It is to be noted that the entanglement does not depend on the value of $\chi'-2\delta$ [see Eqs.~(\ref{vis-dr:a})--(\ref{vdvr-combn})]. 
\par
The state $\widehat{\rho}_1$ is chosen to be fully mixed. (Note that in this case $\eta=0$ and values of $\mathscr{I}$ and $I_H$ are irrelevant.) Applying Eq.~(\ref{PPT-vis}), we find that $\dm_1$ is separable as one would expect. The state $\dm_2$ is chosen as a Werner state with $\eta=0.2 < 1/3$ and this state is found to be separable as expected. We choose $\widehat{\rho}_3$ as an entangled mixed state and we find that Eq.~(\ref{PPT-vis}) verifies that it is entangled. The density matrix $\widehat{\rho}_4$  represents a Werner state for which $\eta=0.7 > 1/3$. We find this state to be entangled. Finally, the state $\widehat{\rho}_5$ is chosen to be the Bell state, $\ket{\Phi^+}=(\ket{H_IH_S}+\ket{V_IV_S})/\sqrt{2}$. As expected, the state is entangled according to Eq.~(\ref{PPT-vis}). 
\par
In order to express the concurrence in terms of experimentally measurable quantities, we substitute from Eqs.~(\ref{vdvr-combn}) and (\ref{eta}) into Eq.~(\ref{conc-form}) and find that
\begin{align}\label{conc-visbltis}
C(\widehat{\rho})=\text{max}\Big\{0,\frac{\mathcal{V}_{HV}+4\mathcal{V}_{DR}-1}{2}\Big\},
\end{align}
where $\mathcal{V}_{HV}$ and $\mathcal{V}_{DR}$ represent right-hand sides of Eqs.~(\ref{eta}) and (\ref{vdvr-combn}), respectively. Equation (\ref{conc-visbltis}) shows that the concurrence of the two-qubit mixed state can be determined from visibilities of the interference patterns obtained by detecting only one of the qubits.
\begin{figure}[htbp]  \centering
	\includegraphics[width=0.75\linewidth]{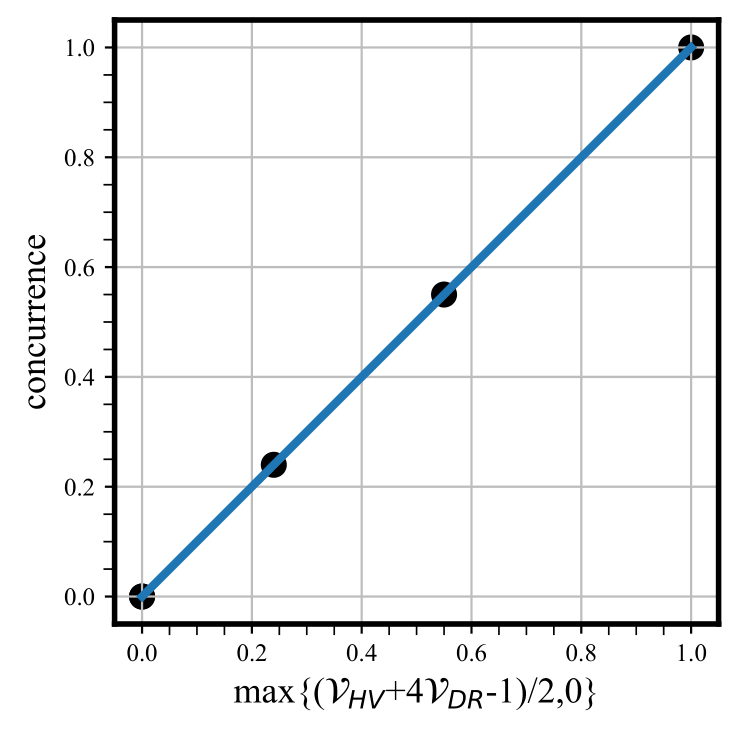}
	\qquad
	\caption{ Concurrence of five generalized Werner states (Table \ref{tab:1}) determined using single-photon visibilities. Simulated data points (filled circles) represent the concurrence computed using state-parameters [Eq.~(\ref{conc-form})] and the formula $\text{max}\{0,(\mathcal{V}_{HV}+4\mathcal{V}_{DR}-1)/2\}$. (Data points for $\dm_1$ and $\dm_2$ coincide.) All simulated data points lie on the straight line predicted by Eq.~(\ref{conc-visbltis}) showing that the concurrence can be determined from single-photon visibilities.} \label{fig:conc-plot}
\end{figure}  
\par
To illustrate Eq.~(\ref{conc-visbltis}), we consider the five quantum states ($\dm_1,\dots,\dm_5$) given in Table \ref{tab:1} and choose $|b_1|= |b_2|=1/\sqrt{2}$ for simplicity. We determine the concurrence of these states using two methods: (i) using the state-parameters [i.e., using Eq.~(\ref{conc-form}); the corresponding  values of the concurrence are given in Table \ref{tab:1}], and (ii) using the values of visibilities [i.e, the formula $\text{max}\{0,(\mathcal{V}_{HV}+2\mathcal{V}_{DR}-1)/2\}$]. In Fig.\,\ref{fig:conc-plot} we plot the values of concurrence obtained by method (i) against the values obtained by method (ii), and find that they lie on the straight line predicted by Eq.~(\ref{conc-visbltis}).
\par
We have thus shown that it is possible to test the PPT criterion and measure concurrence for a two-qubit generalized Werner state without detecting one of the qubits. 
\par
We have expressed the PPT criterion and the concurrence in terms of interference visibilities to be consistent with existing results \cite{pol-ent-theory}. We note that they can be equivalently expressed in terms of detection probabilities [see Appendix E, Eqs.~(\ref{PPT-prob}) and (\ref{conc-prob})]. The latter approach results in simpler theoretical treatment of the problem in the presence of experimental losses (Supplementary Material).
\par
We stress that all state parameters ($\eta$, $I_H$, and $\mathscr{I}$) are not required to be determined for verifying and measuring entanglement, albeit they can be determined from the single-photon interference patterns. Equation (\ref{eta}) [or, equivalently, Eq.~(\ref{P_HV}) of Appendix E] provides an expression for $\eta$. Expressions for $I_H$ and $\mathscr{I}$ are given by Eqs.~(\ref{i-h}) and (\ref{i}) in Appendix F.

\section{Comparison with Existing Results}\label{sec:comp}
In order to put our work into context with existing work, we now compare our results to those presented in Refs.\,\cite{pol-ent-theory,pol-ent-exp}. The entanglement of quantum states considered in Refs.\,\cite{pol-ent-theory,pol-ent-exp} is fully characterized by one coherence term (off-diagonal element) of the density matrix. Consequently, such states are entangled if $\mathcal{V}_D|_{\theta=\frac{\pi}{4}} \neq 0$ or $\mathcal{V}_R|_{\theta=\frac{\pi}{4}} \neq 0$ and vice versa \cite{pol-ent-theory,pol-ent-exp}. This is no longer the case for generalized Werner states as illustrated by state $\dm_2$ in Table \ref{tab:1}. State $\dm_2$ is characterized by state parameters $\eta=0.2$, $\I=1$, and $I_H=0.5$. We find that this state is separable (not entangled) even if $\mathcal{V}_D|_{\theta=\frac{\pi}{4}} \neq 0$ and $\mathcal{V}_R|_{\theta=\frac{\pi}{4}} \neq 0$. Furthermore, for the states considered in Refs.\,\cite{pol-ent-theory,pol-ent-exp}, one does not require to measure visibilities for $H$ and $V$ polarized signal photons to characterize entanglement in a loss-less scenario: measurements in the $H/V$ basis are required only to account for experimental losses. On the contrary, our results show that measurements in the $H/V$ basis are absolutely essential for characterizing entanglement of a generalized Werner state even in the absence of experimental losses. 

\section{Treating Experimental Imperfections}\label{sec:exp-losses}
In an experiment, numerous imperfections may appear. However, results of Ref.\,\cite{pol-ent-exp} show that most dominant ones are the misalignment of idler beams and loss of idler photons between the two sources. These imperfections need separate attention because they result in the loss of interference-visibility that cannot be compensated in any way.
\par
In the Supplementary Material, we provide a detailed analysis showing how to treat such experimental imperfections. We represent the PPT criterion and the concurrence in terms of experimentally measurable quantities by taking these imperfections into account. Equations (\ref{PPT-prob-loss}) and (\ref{conc-prob-loss}) of Appendix G display the corresponding formulas. Considering the five quantum states given by Table\,\ref{tab:1}, we test the PPT criterion and determine their concurrence in the presence of high experimental loss (Supplementary Material, Table S1) and find that the results are identical to those presented in Table\,\ref{tab:1} and Fig.~\ref{fig:conc-plot}. Our analysis thus shows that the proposed method is experimentally implementable and resistant to experimental losses. 

\section{Summary and Conclusions}\label{sec:summary}
It is a common perception that one must detect both photons to measure entanglement of a two-photon \emph{mixed state}. Only recently it has been demonstrated that by the application of ``quantum indistinguishability by path identity'' \cite{Lahiri2022PI_RMP} it is possible to measure entanglement in a special class of two-qubit mixed states without detecting one of the qubits \cite{pol-ent-theory,pol-ent-exp}. We have generalized the method to cover a wider class of states. In particular, we have shown how to measure entanglement in a two-qubit generalized Werner state without detecting one of the qubits and without employing coincidence measurement or postselection. Our analysis shows that the generalization requires non-trivial adaptation of the previously introduced measurement procedure, especially when experimental imperfections are present. Our work also marks an important step toward the generalization of this unique method to an arbitrary two-qubit mixed state.  
\par
In a recent publication \cite{Zhan2021EntPiHd}, Zhan has proposed to extend the measurement procedure introduced in Refs.\,\cite{pol-ent-theory,pol-ent-exp} to mixed high-dimensional Bell states. We expect that our work will inspire novel proposals to measure entanglement in high-dimensional generalized Werner states, which would not require detection of all entangled particles. We hope that such a generalization will contribute toward reducing the resource requirement for entanglement measurement of high-dimensional entangled states. 
\par
Finally, since our analysis is based on quantum field theory, it can in principle be applied to non-photonic quantum states. As the method relies on detection of only one of the two entangled particles, it can be practically advantageous when adequate detectors are not available for both entangled particles.

\section*{Acknowledgments}
This material is based upon work supported by the Air Force Office of Scientific Research under award number FA9550-23-1-0216.
\section*{Appendix A : The PPT criterion eigenvalues}
\renewcommand{\theequation}{A\arabic{equation}}
\setcounter{equation}{0}
We obtain a partial transposition of the  density matrix, $\widehat{\rho}$ [Eq.~(\ref{q-state})], by taking the transposition with respect to only one of the photons. It can be readily found that the eigenvalues of the resulting matrix are
\begin{subequations}\label{PPT-evals}
	\begin{align}
	\alpha_{1}&=\frac{1-\eta-4\eta\mathscr{I}\sqrt{I_{H}I_{V}}}{4}, \label{PPT-evals:a} \\
	\alpha_{2}&=\frac{1-\eta+4\eta\mathscr{I}\sqrt{I_{H}I_{V}}}{4}, \label{PPT-evals:b} \\
	\alpha_{3}&=\frac{1+3\eta-4\eta I_{H}}{4}, \label{PPT-evals:c} \\
	\alpha_{4}&=\frac{1-\eta+4\eta I_{H}}{4}. \label{PPT-evals:d}
	\end{align}
\end{subequations}
We show below that only $\alpha_1$ can take negative values and the remaining eigenvalues must be non-negative. 
\par
Since $0\leq I_{H}\leq 1$ and $I_V=1-I_H$, we must have $0\leq \sqrt{I_{H}I_V}\leq 1/2$. Furthermore, since $0\leq \I \leq 1$, we obtain the condition 
\begin{align}\label{PPT-step-1}
0\leq \I\sqrt{I_{H}I_V}\leq 1/2.
\end{align}
Applying condition (\ref{PPT-step-1}) to Eq.~(\ref{PPT-evals:a}), we immediately find that 
\begin{align}\label{neg-eval-ineq}
\frac{1-3\eta}{4}\leq \alpha_1 \leq \frac{1-\eta}{4}.
\end{align}
The expression on the left of inequality (\ref{neg-eval-ineq}), is negative when $\eta > 1/3$, that is, the lower bound of $\alpha_1$ can be negative. For example, if one sets $I_H=1/2$, $\I=1$, and $\eta=2/3$, one finds from Eq.~(\ref{PPT-evals:a}) that $\alpha_1= -1/4$.
\par
We now consider eigenvalue $\alpha_{2}$. We note that $1-\eta \geq 0$ and $4\eta\mathscr{I}\sqrt{I_{H}I_{V}} \geq 0$. It now becomes evident from Eq.~(\ref{PPT-evals:b}) that $\alpha_2\geq0$.
\par
Using Eq.~(\ref{PPT-evals:c}), we can express $\alpha_3$ in the following form
\begin{align}\label{a3-form}
\alpha_{3}=\frac{1-\eta I_H}{4}+\frac{3\eta(1-I_{H})}{4}.
\end{align}
Since $0\leq \eta \leq 1$ and $0\leq I_{H}\leq 1$, it immediately follows that $\alpha_3\geq 0$. 
\par
Finally, since $1-\eta \geq 0$ and $4\eta I_H \geq 0$, we readily obtain from Eq.~(\ref{PPT-evals:d}) that $\alpha_4 \geq 0$.

\section*{Appendix B: Derivation of Eq.~(\ref{conc-form})}
\renewcommand{\theequation}{B\arabic{equation}}
\setcounter{equation}{0}
The concurrence is determined following the prescription given in \cite{wootters1998entanglement}. First, the spin flipped density matrix $\widehat{\widetilde{\rho}}$ is determined using the formula
\begin{align}\label{spin-flip-density}
\widehat{\widetilde{\rho}}=(\widehat{\sigma}_y\otimes\widehat{\sigma}_y)\widehat{\rho}^*(\widehat{\sigma}_y\otimes\widehat{\sigma}_y),  
\end{align}
where $\widehat{\sigma}_y$ is the second Pauli matrix, $\otimes$ represents the Kronecker product, the asterisk $(\ast)$ implies complex conjugation, and $\dm$ is given by Eq.~(\ref{q-state}). It is well known that $\widehat{\widetilde{\rho}}$ has only non-negative eigenvalues \cite{wootters1998entanglement}. We denote these eigenvalues by $\lambda_1^2$, $\lambda_2^2$, $\lambda_3^2$, and $\lambda_4^2$. We find these eigenvalues to be given by 
\begin{subequations}\label{conc-square-evals}
	\begin{align}
	&\lambda_1^2=\lambda_2^2=\frac{(1-\eta)^2}{16}, \label{conc-square-evals1} \\
	& \lambda_3^2=\frac{(1-\eta)^2}{16}+c_1-c_2, \label{conc-square-evals2} \\
	&\lambda_4^2=\frac{(1-\eta)^2}{16}+c_1+c_2, \label{conc-square-evals3} 
	\end{align}
\end{subequations}
where
\begin{subequations}
	\begin{align}
	& c_1=\frac{\eta}{4}\left(1-\eta\right)+\eta^2I_HI_V(1+\mathscr{I}^2), \label{c1} \\
	& c_2=\frac{\eta\mathscr{I}}{2}\sqrt{I_HI_V[1+2\eta-\eta^2(3-16I_HI_V)]}. \label{c2}
	\end{align}
\end{subequations}
It is evident from Eqs.~(\ref{c1}) and (\ref{c2}) that $c_1 \geq 0$ and $c_2 \geq 0$. Therefore, we have from Eqs.~(\ref{conc-square-evals1})--(\ref{conc-square-evals3}) that 
$\lambda_4\geq \lambda_3$ and $\lambda_4\geq \lambda_1=\lambda_2$. Consequently, the standard formula for the concurrence, $C(\widehat{\rho})=\text{max}\{\lambda_4-\lambda_3-\lambda_2-\lambda_1,0\}$ \cite{wootters1998entanglement}, becomes
\begin{align}\label{conc-formula}
C(\widehat{\rho})=\text{max}\{\lambda_4-\lambda_3-2\lambda_1,0\},
\end{align}
where $\lambda_1,\dots,\lambda_4$ are positive square-roots of $\lambda_1^2,\dots,\lambda_4^2$.
\par
We now observe from Eqs.~(\ref{PPT-evals:a}) and (\ref{conc-square-evals}) that the following relation holds:
\begin{align}\label{conc-reltn}
\left[\lambda_4^2+\lambda_3^2-4(\lambda_1-\alpha_1)^2\right]^2=4\lambda_4^2\lambda_3^2.
\end{align}
It immediately follows from Eq.~(\ref{conc-reltn}) that
\begin{align}\label{conc-step}
\lambda_4-\lambda_3-2\lambda_1=-2\alpha_{1}.
\end{align}
Combining Eqs.~(\ref{PPT-evals:a}), (\ref{conc-formula}), and  (\ref{conc-step}) we immediately obtain the concurrence-formula given by Eq.~(\ref{conc-form}).

\section*{Appendix C: The two-photon density matrix}
\renewcommand{\theequation}{C\arabic{equation}}
\setcounter{equation}{0}
We briefly discuss the procedure of obtaining the two-photon density matrix generated by our system. We start with the generalized Werner state [Eq.~(\ref{q-state-2})]:
\begin{align}\label{q-state-2-app}
\dm=\sum_{\mu,\nu}^{H,V} \sum_{\mu',\nu'}^{H,V} \bra{\mu_I\nu_S}\dm\ket{\mu_I'\nu_S'} \ket{\mu_I\nu_S} \bra {\mu_I'\nu_S'}.
\end{align}
Without any loss of generality, an arbitrary element of this density matrix can be expressed in the form
\begin{align}\label{q-state-elem}
\bra{\mu_I\nu_S}\dm\ket{\mu_I'\nu_S'}=C_{\mu\nu} C_{\mu'\nu'} J_{\mu\nu}^{\mu'\nu'} \exp\big( i\phi_{\mu\nu}^{\mu'\nu'}\big),
\end{align}
where each quantity on the right-hand side is defined as follows. The real and non-negative quantity, $C_{\mu\nu}=\sqrt{\bra{\mu_I\nu_S}\dm\ket{\mu_I \nu_S} }$, represents the square-root of a diagonal element of $\dm$. The quantity $J_{\mu\nu}^{\mu'\nu'}$ is also non-negative and given by the properties: (i) $J_{\mu\nu}^{\mu'\nu'} =J_{\mu'\nu'}^{\mu\nu}$, (ii) $J_{\mu\nu}^{\mu'\nu'} =1$ for all diagonal elements, i.e., for $\mu=\mu'$ and $ \nu=\nu'$, (iii) for $\mu\neq\mu'$, $\nu\neq \nu'$, $\mu=\nu$ and $\mu'=\nu'$, it takes the following form:
\begin{align} \label{I-form}
J_{HH}^{VV} =J_{VV}^{HH} = \frac{4\I \eta \sqrt{I_HI_V}}{\sqrt{(1-\eta+4\eta I_H)(1-\eta+4\eta I_V)}},
\end{align}
and (iv) $J_{\mu\nu}^{\mu'\nu'} =0$ for the remaining cases. Phases $\phi_{\mu\nu}^{\mu'\nu'}$ in Eq.~(\ref{q-state-elem}) obey the following relations: $\phi_{\mu\nu}^{\mu'\nu'}=-\phi_{\mu'\nu'}^{\mu\nu}$, $\phi_{HH}^{VV}=-\phi$, and values of $\phi_{\mu\nu}^{\mu'\nu'}$ for other choices of $\mu, \nu, \mu', \nu'$ are irrelevant since the corresponding matrix elements are zero. The quantum state given by Eqs.~(\ref{q-state-2-app})-(\ref{I-form}) is generated by an individual source. 
\par
We first consider the case in which the idler beams are not aligned. In our system, the two sources are mutually coherent and they emit in such a way that there are never more than one photon pair simultaneously. A prescription to write down the quantum state in such a scenario is given in Ref.\,\cite{pol-ent-theory}. Following this prescription, we find that the state jointly generated by the two sources is represented by the $8\times 8$ density matrix
\begin{align}\label{q-st-two-source}
\dm''=\sum_{j,k}^{1,2} b_j^{\ast} b_k \sum_{\mu,\nu}^{H,V} \sum_{\mu',\nu'}^{H,V} &C_{\mu \nu} C^*_{\mu' \nu'} J_{\mu\nu}^{\mu'\nu'} \exp\big( i\phi_{\mu_k\nu_k}^{\mu'_j\nu'_j}\big) \nonumber \\  &\times \ket{\mu_{I_k}\nu_{S_k}} \bra{\mu'_{I_j}\nu'_{S_j}},
\end{align}
where $j=1,2$ and $k=1,2$ label the two sources, $|b_j|^2$ is the probability of emission from source $Q_j$ (i.e., $|b_1|^2+|b_2|^2=1$), quantities $C_{\mu \nu}$ and $J_{\mu\nu}^{\mu'\nu'}$ are defined below Eq.~(\ref{q-state-elem}), $\phi_{\mu_k\nu_k}^{\mu'_j\nu'_j}=-\phi_{\mu'_j\nu'_j}^{\mu_k\nu_k}$, and $\phi_{\mu_j\nu_j}^{\mu'_j\nu'_j}=\phi_{\mu\nu}^{\mu'\nu'}$. 
\par
When the idler beams are aligned and the unitary transformation [Eq.~(\ref{idler-transform})] is performed on the state of the idler photon between $Q_1$ and $Q_2$, the transformation of kets are given by Eqs.~(\ref{align-cond-a}) and (\ref{align-cond-b}). We rewrite these equations in the following form:
\begin{align}\label{align-app}
\ket{\mu_{I_2}}=e^{-i\phi_I} \sum_{\lambda}^{H,V} U_{\mu\lambda}^{\ast} \ket{\lambda_{I_1}},
\end{align}
where $U_{\mu\lambda}$ represents matrix elements of the unitary transformation given by Eq.~(\ref{idler-transform}).
\par	
The two-photon quantum state ($\dm^{(f)}$) generated by our system is obtained by substituting from Eq.~(\ref{align-app}) into Eq.~(\ref{q-st-two-source}) and it is given by
\begin{widetext}
	\begin{align}\label{dm-final-exp}
	&\dm^{(f)}=|b_1|^2\sum_{\mu,\nu}^{H,V}\Big[\eta I_\mu\ket{\mu_{I_1},\mu_{S_1}}\bra{\mu_{I_1},\mu_{S_1}}+\Big(\mathscr{I}\eta\sqrt{I_HI_V}e^{-i\phi}\ket{H_{I_1},H_{S_1}}\bra{V_{I_1},V_{S_1}}+\text{H.C.}\Big)+\frac{1-\eta}{4}\ket{\mu_{I_1},\nu_{S_1}}\bra{\mu_{I_1},\nu_{S_1}}\Big] \nonumber \\ &  +|b_2|^2\sum_{\mu,\nu}^{H,V}\Big[\eta I_\mu \sum_{\lambda,\lambda'}^{H,V} U^*_{\mu\lambda}U_{\mu\lambda'}\ket{\lambda_{I_1},\mu_{S_2}}\bra{\lambda'_{I1},\mu_{S_2}} +\Big(\sum_{\lambda,\lambda'}^{H,V}\mathscr{I}\eta\sqrt{I_HI_V}e^{-i\phi}U^*_{H\lambda} U_{V\lambda'}\ket{\lambda_{I_1},H_{S_2}}\bra{\lambda'_{I1},V_{S_2}}+\text{H.C.}\Big) \nonumber \\ & +\frac{1-\eta}{4}\sum_{\lambda,\lambda'}^{H,V}U^*_{\mu\lambda}U_{\mu\lambda'}\ket{\lambda_{I_1},\nu_{S_2}}\bra{\lambda'_{I1},\nu_{S_2}} \Big] +\Big\{ e^{i\phi_I}b_1b_2^* \Big[\sum_{\mu,\nu}^{H,V}\Big( \sum_{\lambda}^{H,V}\eta I_\mu e^{i\phi_{\mu_1\mu_1}^{\mu_2\mu_2}} U_{\mu\lambda} \ket{\mu_{I_1},\mu_{S_1}} \bra{\lambda_{I_1},\mu_{S_2}} \nonumber \\ & +\frac{1-\eta}{4}\sum_{\lambda}^{H,V} e^{i\phi_{\mu_1\nu_1}^{\mu_2\nu_2}} U_{\mu\lambda}\ket{\mu_{I_1},\nu_{S_1}} \bra{\lambda_{I_1},\nu_{S_2}} \Big) +\mathscr{I}\eta \sqrt{I_HI_V}\sum_{\mu\neq \nu}^{H,V} \sum_{\lambda}^{H,V}  e^{i\phi_{\mu_1\mu_1}^{\nu_2 \nu_2}} U_{\nu\lambda}\ket{\mu_{I_1},\mu_{S_1}}\bra{\lambda_{I_1},\nu_{S_2}}\Big]+\text{H.C.}\Big\}
	\end{align}
\end{widetext}

\section*{Appendix D: General expressions for signal photon detection probabilities (photon counting rates)}
\renewcommand{\theequation}{D\arabic{equation}}
\setcounter{equation}{0}
We obtain the detection probability, $P_{\mu}$, of a signal photon with polarization $\mu$ (where $\mu=H,V,D,A,R,L$) at an output of $BS$ (Fig.~\ref{fig:scheme}) using Eqs.~(\ref{signal-dm})-(\ref{ph-count-rt}). When the signal photon is projected onto $\ket{H_S}$ polarization state, we have from these equations
\begin{align}\label{rh-first}
P_{H}&=\frac{1-\eta}{4}+\frac{\eta I_{H}}{2}+|b_{1}||b_{2}| \Big\{ \sin (\phi_\text{in}+\phi_{H_1H_1}^{H_2H_2}-\delta) \nonumber \\ & \times \Big(\eta I_{H}+\frac{1-\eta}{4}\Big) -\Big(\frac{1-\eta}{4}\Big) \sin(\phi_\text{in}+\phi_{V_1H_1}^{V_2H_2}+\delta)\Big\} \nonumber \\ & \times \cos(2\theta),
\end{align}
where $\phi_{\text{in}}=\arg\{b_1\}-\arg\{b_2\}+\phi_I-\phi_S$. We now note the following trigonometric identity:
\begin{align}\label{trig:identity}
&u\sin x+v\sin (x+\alpha)\nonumber \\ & =\{u^2+v^2+2uv\cos \alpha \}^{\frac{1}{2}} \sin(x+\beta),
\end{align}
where $\tan \beta=u\sin \alpha/(u+v\cos\alpha)$.
Now using  Eqs.~(\ref{rh-first}) and (\ref{trig:identity}), we find that
\begin{align}\label{ph-countingrate-h-full}
&P_H=\frac{1-\eta}{4}+\frac{\eta I_{H}}{2}+|b_{1}||b_{2}| \cos(2\theta) \Big\{\Big(\eta I_{H}+\frac{1-\eta}{4}\Big)^2 \nonumber \\ &  +\Big(\frac{1-\eta}{4}\Big)^2-2\Big(\eta I_H+\frac{1-\eta}{4}\Big) \Big(\frac{1-\eta}{4}\Big) \cos (\chi+2\delta)\Big\}^{\frac{1}{2}} \nonumber \\ & \times \sin(\phi_{\text{in}}+\phi_0),
\end{align}
where $\chi=\phi^{V_2H_2}_{V_1H_1}-\phi^{H_2H_2}_{H_1H_1}$, $\phi_0=\phi_{H_1H_{1}}^{H_2H_2}-\delta+\epsilon_1$ and $\epsilon_1$ is analogous to $\beta$ in Eq.~(\ref{trig:identity}). It can readily checked that Eq.~(\ref{ph-countingrate-h-full}) reduces to Eq.~(\ref{ph-counting-rates-h}) when $\theta=0$.
\par
Following the same procedure, we determine the detection probability (i.e., the single-photon counting rate) when the signal photon is projected onto the $\ket{V_S}$ polarization state. We find it to be given by
\begin{align}\label{ph-countingrate-v-full}
&P_V=\frac{1-\eta}{4}+\frac{\eta I_{V}}{2}+|b_{1}||b_{2}| \cos(2\theta) \Big\{\Big(\eta I_{V}+\frac{1-\eta}{4}\Big)^2 \nonumber \\ &+\Big(\frac{1-\eta}{4}\Big)^2-2\Big(\eta I_V+\frac{1-\eta}{4}\Big)\Big(\frac{1-\eta}{4} \Big) \cos (\chi''+2\delta)\Big\}^{\frac{1}{2}}  \nonumber \\ & \times \sin(\phi_{\text{in}}+\phi_1),
\end{align}
where $\chi''=\phi^{V_2V_2}_{V_1V_1}-\phi^{H_2V_2}_{H_1V_1}$, $\phi_1=\phi_{H_1V_1}^{H_2V_2}-\delta+\epsilon_2$, and $\epsilon_2$ is analogous to $\beta$ in Eq.~(\ref{trig:identity}).
\par
Likewise, using Eqs.~(\ref{signal-dm})-(\ref{ph-count-rt}), we determine the photon counting rates corresponding to the polarization states $\ket{D_S}$, $\ket{A_S}$, $\ket{R_S}$ and $\ket{L_S}$ 
\begin{subequations}\label{ph-coutng-rate-DARL}
	\begin{align}
	&P_D=\frac{1}{4}+\frac{|b_1||b_2|}{2}\big[ \eta\mathscr{I}\sqrt{I_HI_V}  \sqrt{2+2\cos (\chi'-2\delta)} \nonumber\\ &  \times \sin (\phi_{\text{in}} + \phi_{H_1H_{1}}^{V_2V_{2}}+\delta +\epsilon_3) \sin 2\theta +\mathcal{W} \cos 2\theta \big],  \label{ph-coutng-rate-DARL:a}\\
	&P_A=\frac{1}{4}+\frac{|b_1||b_2|}{2}\big[ - \eta\mathscr{I}\sqrt{I_HI_V}  \sqrt{2+2\cos (\chi'-2\delta)} \nonumber\\ & \times \sin (\phi_{\text{in}} + \phi_{H_1H_{1}}^{V_2V_{2}}+\delta +\epsilon_3) \sin 2\theta +\mathcal{W} \cos 2\theta \big], \label{ph-coutng-rate-DARL:b}\\ 
	&P_R=\frac{1}{4}+\frac{|b_1||b_2|}{2}\big[- \eta\mathscr{I}\sqrt{I_HI_V}  \sqrt{2-2\cos (\chi'-2\delta)} \nonumber\\ & \times \cos (\phi_{\text{in}} + \phi_{H_1H_{1}}^{V_2V_{2}}+\delta +\epsilon_4) \sin 2\theta +\mathcal{W} \cos 2\theta \big], \label{ph-coutng-rate-DARL:c}\\
	&P_L=\frac{1}{4}+\frac{|b_1||b_2|}{2}\big[ \eta\mathscr{I}\sqrt{I_HI_V}  \sqrt{2-2\cos (\chi'-2\delta)} \nonumber\\ & \times \cos (\phi_{\text{in}} + \phi_{H_1H_{1}}^{V_2V_{2}}+\delta +\epsilon_4) \sin 2\theta +\mathcal{W} \cos 2\theta \big], \label{ph-coutng-rate-DARL:d}
	\end{align}
\end{subequations}
where $\phi_{\text{in}}=\arg\{b_1\}-\arg\{b_2\}+\phi_I-\phi_S$, $\chi'=\phi_{V_1V_1}^{H_2H_2}-\phi_{H_1H_1}^{V_2V_2}$, $\epsilon_3$ and $\epsilon_4$ are analogous to $\beta$ in Eq.~(\ref{trig:identity}), and 
\begin{align}\label{W-form}
&\mathcal{W}=\Big(\eta I_{H}+\frac{1-\eta}{4}\Big) \sin (\phi_{\text{in}}+\phi_{H_1H_{1}}^{H_2H_{2}}-\delta) \nonumber \\ & -\frac{1-\eta}{4} [\sin (\phi_{\text{in}}+\phi_{V_1H_{1}}^{V_2H_{2}}+\delta)-\sin (\phi_{\text{in}}+\phi_{H_1V_{1}}^{H_2V_{2}}-\delta)] \nonumber \\ & -\Big(\eta I_{V}+\frac{1-\eta}{4}\Big) \sin (\phi_{\text{in}}+\phi_{V_1V_{1}}^{V_2V_{2}}+\delta).
\end{align}
expressions for visibility for $D$, $A$, $R$, and $L$ polarizations [Eqs.~(\ref{vis-dr:a}) and (\ref{vis-dr:b})] are obtained by setting $\theta=\pi/4$ in Eqs.~(\ref{ph-coutng-rate-DARL:a})-(\ref{ph-coutng-rate-DARL:d}), followed by application of the standard formula for visibility.

\section*{Appendix E: Alternative Expressions for the PPT Criterion and Concurrence}
\renewcommand{\theequation}{E\arabic{equation}}
\setcounter{equation}{0}
Here we express the PPT Criterion and Concurrence in terms of single-photon detection probabilities. In Eqs.~(\ref{ph-countingrate-h-full}) and (\ref{ph-countingrate-v-full}), we set $\theta=0$ to maximize the contribution from the interference term and choose $\delta_H$  and $\delta_V$ such that $\cos(\chi+2\delta_H)=1$ and $\cos(\chi''+2\delta_V)=1$.  Let us define
\begin{subequations}\label{PTM}
	\begin{align}
	P_\mu^{(+)}&=\left(P_\mu\big|_{\theta=0}^{\delta=\delta_\mu}\right)_{\text{max}}+\left(P_\mu\big|_{\theta=0}^{\delta=\delta_\mu}\right)_{\text{min}}, \label{PTM:a}\\
	P_\mu^{(-)}&=\big(P_\mu\big|_{\theta=0}^{\delta=\delta_\mu}\big)_{\text{max}}-\big(P_\mu\big|_{\theta=0}^{\delta=\delta_\mu}\big)_{\text{min}}\label{PTM:b},
	\end{align}
\end{subequations}
where $\mu=H,V$ and the maximum and minimum values of $P_\mu\big|_{\theta=0}^{\delta=\delta_\mu}$ are obtained by varying the interferometric phase $\phi_{\text{in}}$. It readily follows from Eqs.~(\ref{ph-countingrate-h-full}), (\ref{ph-countingrate-v-full}), (\ref{PTM:a}) and (\ref{PTM:b}) that
\begin{align}\label{P_HV}
\eta=\frac{P_H^{(-)}}{|b_1||b_2|}-2P_H^{(+)}+1,
\end{align}
where $H$ can be replaced by $V$ and an expression for $|b_1||b_2|$ in terms of single-photon detection probabilities is given by Eq.~(\ref{b1b2}).
\par
Likewise, for $D$, $A$, $R$, and $L$ polarizations, we define
\begin{align}\label{PM-DARL}
P_{\nu}^{(-)}=\big(P_{\nu}\big|_{\theta=\frac{\pi}{4}}\big)_{\text{max}}-\big(P_{\nu}\big|_{\theta=\frac{\pi}{4}}\big)_{\text{min}},
\end{align}
where $\nu=D,A,R,L$. Using Eqs.~(\ref{ph-coutng-rate-DARL:a})-(\ref{ph-coutng-rate-DARL:d}) and Eq.~(\ref{PM-DARL}), we find that 
\begin{align}
\eta\I\sqrt{I_HI_V}=\frac{1}{2|b_1||b_2|}\sqrt{\big[P_D^{(-)}\big]^2+\big[P_R^{(-)}\big]^2}, \label{P_DR}
\end{align}
where $|b_1||b_2|$ is given by Eq.~(\ref{b1b2}). From Eqs.~(\ref{PPT-eval}), (\ref{P_HV}), and (\ref{P_DR}), we have 
\begin{align}
\alpha_1=\frac{1-\widetilde{P}_H -4\widebar{P}_{DR}}{4},
\end{align}
where we have denoted the right-hand sides of Eqs.~(\ref{P_HV}) and (\ref{P_DR}) by $\widetilde{P}_H$ and $\widebar{P}_{DR}$ respectively.
According to the PPT criterion, the state $\dm$ [Eq.~(\ref{q-state})] is entangled if and only if $\alpha_1<0$, that is, if and only if
\begin{align}\label{PPT-prob}
\widetilde{P}_H+4\widebar{P}_{DR} >1,
\end{align}
where $H$, $D$, and $R$ can be replaced by $V$, $A$, and $L$ respectively.
\par
In order to express the concurrence in terms of single-photon detection probabilities, we substitute from Eqs.~(\ref{P_DR}) and (\ref{P_HV}) into Eq.~(\ref{conc-form}) and find that
\begin{align}\label{conc-prob}
C(\widehat{\rho})=\text{max}\Big\{0,\frac{\widetilde{P}_H+4\widebar{P}_{DR}-1}{2}\Big\},
\end{align}
where $H$, $D$, and $R$ can be replaced by $V$, $A$, and $L$ respectively.
\par
We have thus represented the PPT criterion and concurrence in terms of single-photon detection probabilities. We stress that Eqs.~(\ref{PPT-prob}) and (\ref{conc-prob}) are equivalent to Eqs.~(\ref{PPT-vis}) and (\ref{conc-visbltis}) respectively.

\section*{Appendix F: Expressions for $I_H$ and $\I$}
\renewcommand{\theequation}{F\arabic{equation}}
\setcounter{equation}{0}
In this appendix, we express state-parameters $I_H$ and $\I$ in terms of experimentally measurable quantities. Using Eqs.~(\ref{ph-countingrate-h-full}), (\ref{PTM:a}), and (\ref{PTM:b}), we find that
\begin{align}\label{i-h}
I_H=\frac{P_H^{(-)}}{2P_H^{(-)}+|b_1||b_2|\left(2-4P_H^{(+)}\right)},
\end{align}
where an expression for $|b_1||b_2|$ in terms of experimentally measurable quantities is given by Eq.~(\ref{b1b2}).
\par
Using Eqs.~(\ref{P_HV}) and (\ref{P_DR}), we find that
\begin{align} \label{i}
\I=\frac{\widebar{P}_{DR}}{\widetilde{P}_H\sqrt{I_H(1-I_H)}},  
\end{align}
where $\widetilde{P}_H$ and $\widebar{P}_{DR}$ are right-hand sides of Eqs.~(\ref{P_HV}) and (\ref{P_DR}) respectively and we have used the relation $I_V=1-I_H$. Now substituting from Eq.~(\ref{i-h}) into Eq.~(\ref{i}), we can obtain an expression for $\mathscr{I}$ in terms of experimentally measurable quantities. Equation (\ref{i}) shows that when $\widetilde{P}_H=0$ (i.e., $\eta=0$) and/or $I_H=0,1$, no meaningful value of $\mathscr{I}$ can be obtained. This is because in these cases all off-diagonal terms of the density matrix [Eq.~(\ref{q-state})] are zero for all values of $\mathscr{I}$, that is, obtaining a value for $\mathscr{I}$ becomes irrelevant in these cases. 

\section*{Appendix G: Effects of Experimental Imperfections}
\renewcommand{\theequation}{G\arabic{equation}}
\setcounter{equation}{0}
Both the misalignment of idler beams and polarization-dependent loss of idler photons between the two sources can be effectively modeled by the action of an attenuator (e.g., neutral density filter or any absorptive plate) that has polarization dependent transmissivity. Suppose that the amplitude transmission coefficient of the attenuator for horizontal ($H$) and vertical ($V$) polarizations are $T_H$ and $T_V$, respectively. Without any loss of generality it can be assumed that $T_H$ and $T_V$ are real quantities obeying relations $0\leq T_H \leq 1$ and $0\leq T_V \leq 1$. A detailed analysis of the problem considering experimental imperfections is given in the supplementary material. Our analysis shows that both $T_H$ and $T_V$ can be determined from experimental data. Therefore, we treat $T_H$ and $T_V$ as experimentally measurable quantities.
\par
We show in the Supplementary Material that
\begin{equation}\label{eta-loss-main}
\eta =
\begin{cases}
\frac{2P_H^{(-)}+|b_1||b_2|\left(T_{H}+T_{V}-4P_H^{(+)}T_{H}\right)}{|b_1||b_2|(T_{V}+T_{H})}& \text{if $T_H \geq T_V$},\\
\frac{2P_V^{(-)}+|b_1||b_2|\left(T_{V}+T_H-4P_V^{(+)}T_{V}\right)}{|b_1||b_2|(T_{H}+T_{V})} & \text{if $T_H \leq T_V$},
\end{cases}       
\end{equation}
\par 
and
\begin{align}\label{P-DR-main}
\eta\I\sqrt{I_HI_V}=\frac{\sqrt{2\big[(P_D^{(-)})^2+(P_R^{(-)})^2\big]}}{2|b_1||b_2|\sqrt{T_H^2+T_V^2}}.
\end{align}
Here $P_H^{(\pm)}$, $P_V^{(\pm)}$, $P_D^{(-)}$, and $P_R^{(-)}$ are the same physical quantities introduced in Appendix E. However, their analytical expressions now involve  $T_H$ and $T_V$ (Supplementary Material).
\par
Now using Eq.~(\ref{PPT-eval}), (\ref{eta-loss-main}), and (\ref{P-DR-main}), we get
\begin{align}\label{alpha-loss}
\alpha_1=\frac{1-\widetilde{P}_{HV}(T_H,T_V)-4\widebar{P}_{DR}}{4},
\end{align}
where $\widetilde{P}_{HV}$ and $\widebar{P}_{DR}$ are right-hand sides of Eqs.~(\ref{eta-loss-main}) and (\ref{P-DR-main}) respectively.
According to the PPT criterion, the state $\dm$ [Eq.~(\ref{q-state})] is entangled if and only if $\alpha_1<0$. Thus the PPT criterion in the presence of experimental imperfections can be expressed as 
\begin{align}\label{PPT-prob-loss}
\widetilde{P}_{HV}(T_H,T_V)+4\widebar{P}_{DR}>1.
\end{align}
\par
Using Eqs.~(\ref{conc-form}), (\ref{eta-loss-main}), and (\ref{P-DR-main}), we get the following expression for the concurrence:
\begin{align}\label{conc-prob-loss}
C(\widehat{\rho})=\text{max}\Big\{0,\frac{\widetilde{P}_{HV}(T_H,T_V)+4\widebar{P}_{DR}-1}{2}\Big\},
\end{align}
where $\widetilde{P}_{HV}$ and $\widebar{P}_{DR}$ are right-hand sides of Eqs.~(\ref{eta-loss-main}) and (\ref{P-DR-main}) respectively.
\par
It can be readily checked that Eqs.~(\ref{PPT-prob-loss}) and (\ref{conc-prob-loss}) reduce to Eqs.~(\ref{PPT-prob}) and (\ref{conc-prob}) when there is no experimental imperfection, i.e., when $T_H=T_V=1$.

	\bibliography{werner.bib}

\clearpage
	
	\section*{Supplementary Material: Entanglement Measurement in the Presence of Dominant Experimental Imperfections}
	\renewcommand{\theequation}{S\arabic{equation}}
	\renewcommand{\thefigure}{S\arabic{figure}}
	\renewcommand{\thetable}{S\arabic{table}}
	\noindent
\textbf{Synopsis}. Although numerous imperfections may appear in an experimental scenario, the experiment reported in Ref.\,\cite{pol-ent-exp} shows that most dominant ones are the misalignment of idler beams and polarization-dependent loss of idler photons between the two sources. Here, we take these experimental imperfections into account to show that the proposed method is resistant to experimental losses and imperfections. In particular, we represent the PPT criterion and the concurrence in terms of experimentally measurable quantities by taking these imperfections into account. We numerically illustrate the results by considering the five quantum states given  by Table I of the main text. \\
\par
\noindent
\textbf{Theoretical Analysis} \\[3 pt]
\emph{Obtaining an Expression of the Density Matrix.}\textemdash ~Since the detailed description of the setup is given in the main paper, we skip it here for brevity. Instead, we focus on treating the key experimental imperfections quantitatively.
\begin{figure}[b]  \centering
	\includegraphics[width=0.98\linewidth]{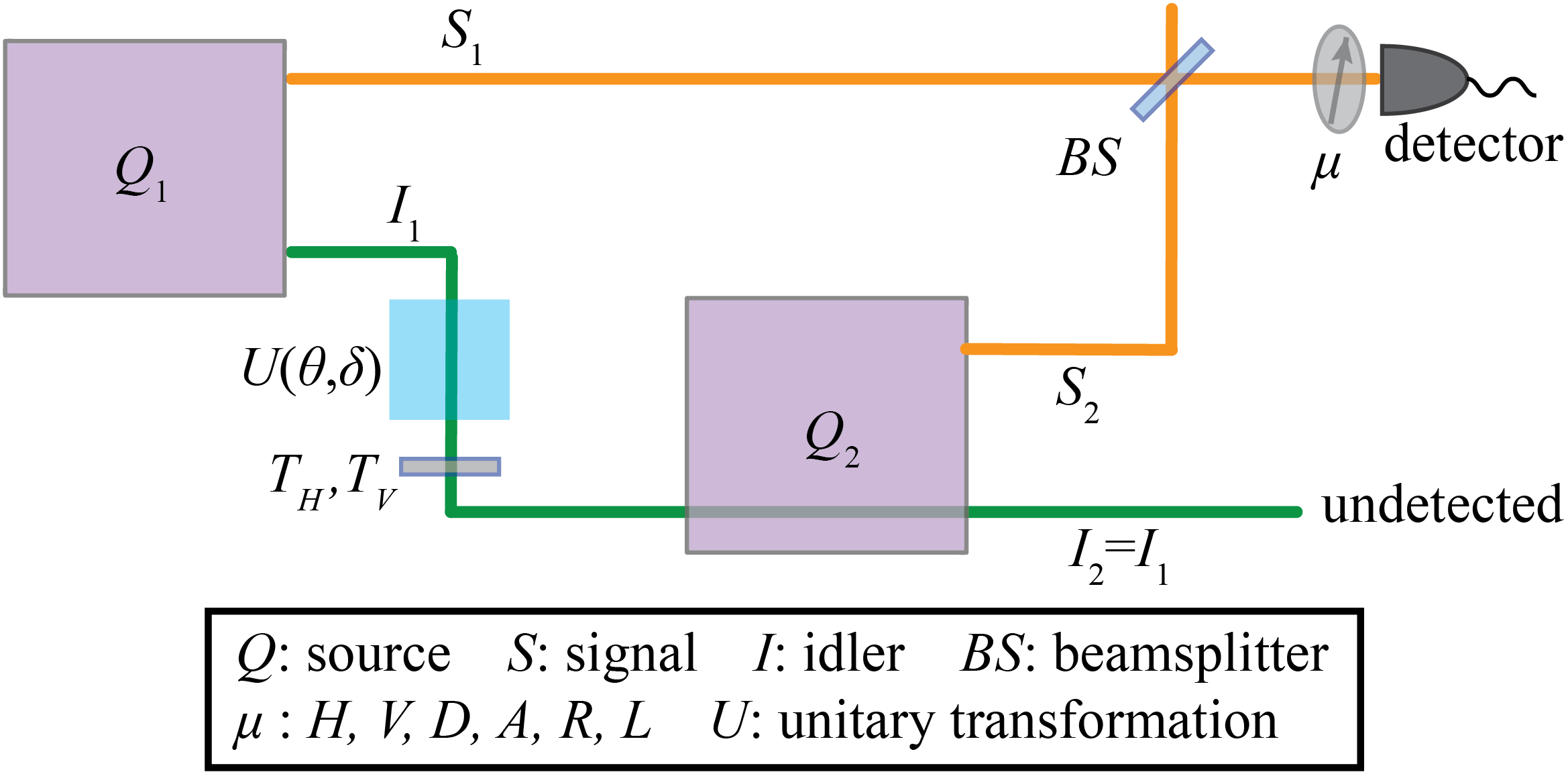}
	\qquad
	\caption{Illustration of the entanglement measurement scheme in presence of experimental imperfections. The misalignment of idler beams and loss of idler photons between the two sources ($Q_1$ and $Q_2$) is modeled by an attenuator with polarization dependent transmissivity. $T_H$ and $T_V$ are amplitude transmission coefficients of the attenuator corresponding to $H$ and $V$ polarizations. Remaining symbols are same as those in Fig.~2 of the main paper.} \label{fig:scheme-supp}
\end{figure} 
\par
We recall from the main paper that the two-qubit generalized Werner state takes the following matrix form in the computational basis $\{ \ket{H_IH_S}, \ket{H_IV_S}, \ket{V_IH_S}, \ket{V_IV_S} \}$: 
\begin{align}\label{q-state-supp}
\dm=\begin{pmatrix}
\eta I_H+\frac{1-\eta}{4} & 0 & 0 & \eta \mathscr{I}\sqrt{I_HI_V}e^{-i\phi} \\
0 & \frac{1-\eta}{4} & 0 & 0\\
0 & 0 & \frac{1-\eta}{4} & 0\\
\eta \mathscr{I}\sqrt{I_HI_V}e^{i\phi} & 0 & 0 & \eta I_V+\frac{1-\eta}{4}
\end{pmatrix},
\end{align}
where $0\leq I_H \leq 1$, $I_V=1-I_H$, $0\leq \eta \leq 1$, $0\leq \I \leq 1$, and $\phi$ represents a phase. This state is generated by an individual source used in the setup. When both sources are considered together and the idler beams are not aligned, the quantum state produced by our system is given by Eq.~(C4) in Appendix C of the main paper. 
\par
We also recall from the main paper that the unitary transformation performed on the idler photon between the two sources has the following form in the $\{ \ket{H}, \ket{V}\}$ basis:
\begin{align}\label{idler-transform-supp}
U(\theta,\delta)=\begin{pmatrix}
e^{-i\delta}\cos 2\theta & e^{-i\delta}\sin 2\theta \\
e^{i\delta}\sin 2\theta & -e^{i\delta}\cos 2\theta 
\end{pmatrix},
\end{align} 
where $0\leq \theta \leq \pi$ and $0\leq \delta/2 \leq \pi$ can be understood as two half-wave plate angles. Calculations up to this step are identical to those in the case with no experimental imperfections. 
\par
As mentioned  above, the two main imperfections are: (i) misalignment of the idler beams and (ii) polarization-dependent loss of idler photons between the two sources. Both of these imperfections can be effectively modeled by the action of an attenuator (e.g., neutral density filter or any absorptive plate) that has polarization dependent transmissivity (Fig.\,\ref{fig:scheme-supp}). Suppose that the amplitude transmission coefficient of the attenuator for horizontal ($H$) and vertical ($V$) polarizations are $T_H$ and $T_V$, respectively. Without any loss of generality it can be assumed that $T_H$ and $T_V$ are real quantities obeying relations $0\leq T_H \leq 1$ and $0\leq T_V \leq 1$. It was shown in Ref.\,\cite{zou1991induced} that the action of an attenuator on the idler-field is mathematically equivalent to that of a beamsplitter with a single input. Therefore, the imperfections in idler beam alignment together with the unitary transformation [Eq.~(\ref{idler-transform-supp})] result in the following transformation of the quantum field representing an idler photon (see also \cite{pol-ent-theory}):
\begin{subequations}\label{align-cond-field-supp}
	\begin{align}
	\opa_{H_{I_2}}&=e^{i\phi_I} \big[T_H e^{-i\delta}\big(\opa_{H_{I_1}}\cos 2\theta + \opa_{V_{I_1}} \sin 2\theta \big) \big] \nonumber \\ & \qquad \quad+ R_H\opa_{H_0}, \label{align-cond-field-supp:a} \\
	\opa_{V_{I_2}}&=e^{i\phi_I} \big[T_V e^{i\delta}\big(\opa_{H_{I_1}}\sin 2\theta - \opa_{V_{I_1}} \cos 2\theta \big) \big] \nonumber \\ & \qquad \quad+ R_V\opa_{V_0}, \label{align-cond-field-supp:b}
	\end{align} 
\end{subequations}
where $\opa$ represents a photon annihilation operator such that $\opa_{\mu_{I_j}}^{\dag}\ket{\text{vacuum}}=\ket{\mu_{I_j}}$ with $\mu=H,V$ and $j=1,2$, the operator $\opa_{\mu_{0}}$ can be interpreted as the field of a lost photon, $R_{\mu}=\sqrt{1-T_{\mu}^2}$, and $\phi_I$ is the phase change due to propagation from $Q_1$ to $Q_2$. Equations (\ref{align-cond-field-supp:a}) and (\ref{align-cond-field-supp:b}) result in the following transformation of kets:
\begin{subequations}\label{align-cond-kets-supp}
	\begin{align}
	\ket{H_{I_2}}&=e^{-i\phi_I} \big[T_H e^{i\delta}\big(\ket{H_{I_1}}\cos 2\theta + \ket{V_{I_1}} \sin 2\theta \big) \big] \nonumber \\ & \qquad \quad+ R_H\ket{H_0}, \label{align-cond-kets-supp:a}\\
	\ket{V_{I_2}}&=e^{-i\phi_I} \big[T_V e^{-i\delta}\big(\ket{H_{I_1}}\sin 2\theta - \ket{V_{I_1}} \cos 2\theta \big) \big] \nonumber \\ & \qquad \quad+ R_V\ket{V_0} \label{align-cond-kets-supp:b},
	\end{align} 
\end{subequations}
where $\ket{\mu_0}=\opa_{\mu_{0}}^{\dag}\ket{\text{vacuum}}$ represents the absorbed photon with polarization $\mu$. It can be readily checked that Eqs.~(\ref{align-cond-kets-supp:a}) and (\ref{align-cond-kets-supp:b}) reduce to Eqs.~(6a) and (6b), respectively, of the main paper when $T_H=T_V=1$ and $R_H=R_V=0$, i.e., when there is no experimental imperfections.
\par
In order to obtain the quantum state, $\dm^{(f)}$, generated by our system (before reaching the beamsplitter $BS$) we substitute from Eqs.~(\ref{align-cond-kets-supp:a}) and (\ref{align-cond-kets-supp:b}) into Eq.~(C4) in Appendix C of the main paper (i.e., the density matrix in the case with unaligned idler beams) and we find that
\begin{widetext}
	\begin{align}\label{dm-final-exp-supp}
	&\dm^{(f)}=|b_1|^2\sum_{\mu,\nu}^{H,V}\Big[\eta I_\mu\ket{\mu_{I_1},\mu_{S_1}}\bra{\mu_{I_1},\mu_{S_1}}+\Big(\mathscr{I}\eta\sqrt{I_HI_V}e^{-i\phi}\ket{H_{I_1},H_{S_1}}\bra{V_{I_1},V_{S_1}}+\text{H.C.}\Big) +\frac{1-\eta}{4}\ket{\mu_{I_1},\nu_{S_1}}\bra{\mu_{I_1},\nu_{S_1}}\Big]\nonumber\\& +|b_2|^2\sum_{\mu,\nu}^{H,V}\Big[\eta I_\mu \Big(\sum_{\lambda,\lambda'}^{H,V}U^*_{\mu\lambda}U_{\mu\lambda'}T_\mu^2 \ket{\lambda_{I_1},\mu_{S_2}}\bra{\lambda'_{I_1},\mu_{S_2}}+ \sum_{\lambda'}^{H,V}T_\mu R_\mu \{U_{\mu\lambda'}\ket{\mu_{0},\mu_{S_2}}\bra{\lambda'_{I_1},\mu_{S_2}}+\text{H.C}\}  \nonumber\\& +R_\mu ^2\ket{\mu_{0},\mu_{S_2}}\bra{\mu_{0},\mu_{S_2}}\Big) +\Big(\sum_{\lambda,\lambda'}^{H,V}\mathscr{I}\eta\sqrt{I_HI_V}e^{-i\phi}\Big\{U^*_{H\lambda} U_{V\lambda'}T_H T_V\ket{\lambda_{I_1},H_{S_2}}\bra{\lambda'_{I_1},V_{S_2}}\nonumber\\& +U^*_{H\lambda} T_H R_V\ket{\lambda_{I_1},H_{S_2}}\bra{V_{0},V_{S_2}} + U_{V\lambda'}R_H T_V\ket{H_{0},H_{S_2}}\bra{\lambda'_{I1},V_{S_2}}+R_H R_V\ket{H_{0},H_{S_2}}\bra{V_0,V_{S_2}}\Big\}+\text{H.C.}\Big)\nonumber\\& +\frac{1-\eta}{4}\Big(\sum_{\lambda,\lambda'}^{H,V}U^*_{\mu\lambda}U_{\mu\lambda'}T_\mu^2 \ket{\lambda_{I_1},\nu_{S_2}}\bra{\lambda'_{I_1},\nu_{S_2}}+\sum_{\lambda'}^{H,V}T_\mu R_\mu \{U_{\mu\lambda'}\ket{\mu_{0},\nu_{S_2}}\bra{\lambda'_{I_1},\nu_{S_2}}+\text{H.C}\} +R_\mu ^2\ket{\mu_{0},\nu_{S_2}}\bra{\mu_{0},\nu_{S_2}}\Big)\Big]\nonumber\\&+\Big[e^{i\phi_I}b_1b_2^*\bigg\{\sum_{\mu,\nu}^{H,V}\Big[\sum_{\lambda}^{H,V}\eta I_\mu e^{i\phi_{\mu_1\mu_1}^{\mu_2\mu_2}}\Big(T_\mu U_{\mu\lambda}\ket{\mu_{I_1},\mu_{S_1}}\bra{\lambda_{I_1},\mu_{S_2}}+R_\mu\ket{\mu_{I_1},\mu_{S_1}}\bra{\mu_{0},\mu_{S_2}}\Big) \nonumber \\ & +\frac{1-\eta}{4}\sum_{\lambda}^{H,V}e^{i\phi_{\mu_1\nu_1}^{\mu_2\nu_2}}\Big(T_\mu U_{\mu\lambda}\ket{\mu_{I_1},\nu_{S_1}}\nonumber\bra{\lambda_{I_1},\nu_{S_2}}+R_\mu\ket{\mu_{I_1},\nu_{S_1}}\nonumber\bra{\mu_0,\nu_{S_2}}\Big)\Big] \nonumber \\ & +\mathscr{I}\eta\sqrt{I_HI_V}\sum_{\mu\neq\nu}^{H,V}e^{i\phi_{\mu_1\mu_1}^{\nu_2\nu_2}}\sum_{\lambda}^{H,V}\Big(T_\nu U_{\nu\lambda}\ket{\mu_{I_1},\mu_{S_1}}\bra{\lambda_{I_1},\nu_{S_2}}+R_\nu\ket{\mu_{I_1},\mu_{S_1}}\bra{\nu_0,\nu_{S_2}}\Big)\bigg\}+\text{H.C.}\Big].
	\end{align}
\end{widetext}
It can be verified that when $T_H=T_V=1$ and $R_H=R_V=0$, Eq.~(\ref{dm-final-exp-supp}) reduces to  Eq.~(C6) in Appendix C of the main paper. 
\par
The reduced density matrix, $\dm_S$,  representing a signal photon before reaching the beamsplitter $BS$, is obtained by taking partial trace of $\dm^{(f)}$ over the subspace of the idler photon and the loss modes. The reduced density matrix is found to be
\begin{widetext}
	\begin{align}\label{signal-dm-supp}
	\dm_S &=K(\eta,I_H) \Big( |b_1|^2 \ket{H_{S1}}\bra{H_{S1}} +|b_2|^2 \ket{H_{S2}}\bra{H_{S2}} \Big) +K(\eta,I_V) \Big( |b_1|^2 \ket{V_{S1}}\bra{V_{S1}} +|b_2|^2 \ket{V_{S2}}\bra{V_{S2}} \Big)  \nonumber \\ & +\Big\{ b_1 b_2^{\ast} e^{i\phi_I} \big[ \Big\{ \L(\eta,I_H,\delta,T_H,T_V)\ket{H_{S1}}\bra{H_{S2}} +\L'(\eta,I_V,\delta,T_H,T_V)\ket{V_{S1}}\bra{V_{S2}} \Big\} \cos 2\theta \nonumber \\ & \qquad \qquad \quad + \eta \I \sqrt{I_HI_V} \Big\{ \Phi(\delta) T_V \ket{H_{S1}}\bra{V_{S2}} +\Phi'(\delta) T_H\ket{V_{S1}}\bra{H_{S2}}  \Big\} \sin 2\theta \big] +\text{H.c.} \Big\},
	\end{align}
\end{widetext}
where H.c.~represents the Hermitian conjugation, $K(\eta,I)=\eta I+(1-\eta)/2$, $\Phi(\delta)=\exp[i(\phi^{V_2,V_2}_{H_1,H_1}+\delta)]$, $\Phi'(\delta)=\exp[i(\phi^{H_2,H_2}_{V_1,V_1}-\delta)]$, and 
\begin{subequations}\label{signal-dm-exp-supp}
	\begin{align}
	&\L(\eta,I_H,\delta,T_H,T_V)= M(\eta,I_H) T_H\exp \Big[i \Big(\phi^{H_2,H_2}_{H_1,H_1}-\delta \Big) \Big] \nonumber \\ &-N(\eta)T_V\exp \Big[i \Big(\phi^{V_2,H_2}_{V_1,H_1}+\delta \Big) \Big], \\ &\L'(\eta,I_V,\delta,T_H,T_V)= N(\eta)T_H\exp \Big[ i \Big(\phi^{H_2,V_2}_{H_1,V_1}-\delta \Big) \Big]  \nonumber \\ & -M(\eta,I_V)T_V \exp \Big[ i \Big(\phi^{V_2,V_2}_{V_1,V_1}+\delta \Big) \Big],
	\end{align}
\end{subequations}	
with $M(\eta,I)=(4\eta I+1-\eta)/4$ and $N(\eta)=(1-\eta)/4$. It can once again be checked that Eqs.~(\ref{signal-dm-supp}) and (\ref{signal-dm-exp-supp}) reduce to Eqs.~(7) and (8), respectively, of the main paper when there is no experimental imperfections (i.e., $T_H=T_V=1$ and $R_H=R_V=0$). \\
\par
\noindent 
\emph{Obtaining Detection Probability of a Signal Photon.}\textemdash ~The probability of detecting a signal photon emerging from the beamsplitter (BS) is obtained following the same procedure described in the main paper. In particular, we apply Eqs.~(9) and (10) from the main paper and use Eq.~(\ref{signal-dm-supp}) for the expression for the density matrix.
\par
When the signal photon is projected onto $\ket{H_S}$ polarization state, the probability of its detection is given by 	
\begin{align}\label{ph-countingrate-h-supp}
&P_H=\frac{1-\eta}{4}+\frac{\eta I_{H}}{2}+|b_{1}||b_{2}| \Big\{ \Big[ T_H\Big(\eta I_{H} +\frac{1-\eta}{4}\Big) \Big]^2 \nonumber \\ &  +\Big[T_V\Big(\frac{1-\eta}{4}\Big) \Big]^2-2T_H T_V\Big(\eta I_H+\frac{1-\eta}{4}\Big) \Big(\frac{1-\eta}{4}\Big) \nonumber \\ & \times \cos (\chi+2\delta)\Big\}^{\frac{1}{2}} \cos(2\theta) \sin(\phi_{\text{in}}+\phi_0),
\end{align}
where $\phi_{\text{in}}$=$\arg\{b_1\}-arg\{b_2\}+\phi_I-\phi_S$, $\chi=\phi^{V_2H_2}_{V_1H_1}-\phi^{H_2H_2}_{H_1H_1}$, $\phi_0=\phi_{H_1H_{1}}^{H_2H_2}-\delta+\epsilon_1$ and $\epsilon_1$ is analogous to $\beta$ in Eq.~(D2) in Appendix D of the main paper. It can be readily checked that Eq.~(\ref{ph-countingrate-h-supp}) reduces to Eq.~(D3) of Appendix D in the main paper when $T_H=T_V=1$. 
\par
Similarly, when the signal photon is projected onto $\ket{V_S}$ polarization state, the probability of its detection is given by 
\begin{align}\label{ph-countingrate-v-supp}
&P_V=\frac{1-\eta}{4}+\frac{\eta I_{V}}{2}+|b_{1}||b_{2}| \Big\{T_V^2\Big(\eta I_{V}+\frac{1-\eta}{4}\Big)^2 \nonumber \\ &+T_H^2\Big(\frac{1-\eta}{4}\Big)^2-2T_HT_V\Big(\eta I_V+\frac{1-\eta}{4}\Big)\Big(\frac{1-\eta}{4} \Big)  \nonumber \\ & \times \cos (\chi''+2\delta)\Big\}^{\frac{1}{2}} \cos(2\theta) \sin(\phi_{\text{in}}+\phi_1),
\end{align}
where $\phi_{\text{in}}$ is given below Eq.~(\ref{ph-countingrate-h-supp}), $\chi''=\phi^{V_2V_2}_{V_1V_1}-\phi^{H_2V_2}_{H_1V_1}$, $\phi_1=\phi_{H_1V_1}^{H_2V_2}-\delta+\epsilon_2$, and $\epsilon_2$ is analogous to $\epsilon_1$.
\par
For $D$, $A$, $R$, and $L$ polarizations, we get likewise
\begin{subequations}\label{ph-coutng-rate-DARL-supp}
	\begin{align}
	P_D=&\frac{1}{4}+\frac{|b_1||b_2|}{2}\big[\mathcal{W} \cos 2\theta+ \eta\mathscr{I}\sqrt{I_HI_V} \sin 2\theta  \nonumber\\ &  \times \sqrt{T_H^2+T_V^2+2T_HT_V\cos (\chi'-2\delta)} \nonumber\\ &  \times  \sin (\phi_{\text{in}} + \phi_{H_1H_{1}}^{V_2V_{2}}+\delta +\epsilon_3) \big],  \label{ph-coutng-rate-DARL-supp:a}\\
	P_A=&\frac{1}{4}+\frac{|b_1||b_2|}{2}\big[\mathcal{W} \cos 2\theta - \eta\mathscr{I}\sqrt{I_HI_V} \sin 2\theta \nonumber\\ & \times \sqrt{T_H^2+T_V^2+2T_HT_V\cos (\chi'-2\delta)} \nonumber\\ & \times \sin (\phi_{\text{in}} + \phi_{H_1H_{1}}^{V_2V_{2}}+\delta +\epsilon_3) \big], \label{ph-coutng-rate-DARL-supp:b}\\ 
	P_R=&\frac{1}{4}+\frac{|b_1||b_2|}{2}\big[\mathcal{W}\cos 2\theta -\eta\mathscr{I}\sqrt{I_HI_V} \sin 2\theta \nonumber\\ & \times \sqrt{T_H^2+T_V^2-2T_HT_V\cos (\chi'-2\delta)} \nonumber\\ & \times \cos (\phi_{\text{in}} + \phi_{H_1H_{1}}^{V_2V_{2}}+\delta +\epsilon_4) \big], \label{ph-coutng-rate-DARL-supp:c}\\
	P_L=&\frac{1}{4}+\frac{|b_1||b_2|}{2}\big[\mathcal{W} \cos 2\theta+ \eta\mathscr{I}\sqrt{I_HI_V} \sin 2\theta \nonumber\\ & \times   \sqrt{T_H^2+T_V^2-2T_HT_V\cos (\chi'-2\delta)} \nonumber\\ & \times \cos (\phi_{\text{in}} + \phi_{H_1H_{1}}^{V_2V_{2}}+\delta +\epsilon_4) \big], \label{ph-coutng-rate-DARL-supp:d}
	\end{align}
\end{subequations}
where
\begin{align}\label{W-form-supp}
&\mathcal{W}=\Big(\eta I_{H}+\frac{1-\eta}{4}\Big) T_H \sin (\phi_{\text{in}}+\phi_{H_1H_{1}}^{H_2H_{2}}-\delta) -\frac{1-\eta}{4}  \nonumber \\ & \times [T_V \sin (\phi_{\text{in}}+\phi_{V_1H_{1}}^{V_2H_{2}}+\delta)- T_H \sin (\phi_{\text{in}}+\phi_{H_1V_{1}}^{H_2V_{2}}-\delta)] \nonumber \\ & -\Big(\eta I_{V}+\frac{1-\eta}{4}\Big) T_V \sin (\phi_{\text{in}}+\phi_{V_1V_{1}}^{V_2V_{2}}+\delta).
\end{align}
It can be checked that all the expressions for detection probabilities are fully consistent with those obtained assuming no experimental imperfections in the main paper. 
\par
The term $|b_1||b_2|$ appearing in the expressions for single-photon counting rates can be expressed in terms of experimentally measurable quantities and is given by [Eq.~(17) of the main paper]
\begin{align}\label{b1b2-supp}
|b_1| |b_2| =\frac{\sqrt{P_{\mu}^{(1)}P_{\mu}^{(2)}}}{P_{\mu}^{(1)}+P_{\mu}^{(2)}},
\end{align}
where $\mu=H,V$.
\par
\noindent
\emph{Testing the PPT criterion and Determining the Concurrence.}\textemdash ~We set $\theta=0$ in Eqs.~(\ref{ph-countingrate-h-supp}) and (\ref{ph-countingrate-v-supp}) to maximize the contribution from the interference term. We then choose $\delta=\delta_H$  and $\delta=\delta_V$ for these two equations so that $\cos(\chi+2\delta_H)=1$ and $\cos(\chi''+2\delta_V)=1$, respectively. Let us now define,
\begin{subequations}\label{P-MT-supp}
	\begin{align}
	P_\mu^{(+)}&=\left(P_\mu\big|_{\theta=0}^{\delta=\delta_\mu}\right)_{\text{max}}+\left(P_\mu\big|_{\theta=0}^{\delta=\delta_\mu}\right)_{\text{min}}, \label{P-MT-supp:a} \\
	P_\mu^{(-)}&=\left(P_\mu\big|_{\theta=0}^{\delta=\delta_\mu}\right)_{\text{max}}-\left(P_\mu\big|_{\theta=0}^{\delta=\delta_\mu}\right)_{\text{min}}\label{P-MT-supp:b},
	\end{align}
\end{subequations}
where $\mu=H,V$ and the maximum and minimum values of $P_\mu\big|_{\theta=0}^{\delta=\delta_\mu}$ are obtained by varying the interferometric phase $\phi_{\text{in}}$. Likewise, we can choose $\delta=\delta'_H$  and $\delta=\delta'_V$ so that $\cos(\chi+2\delta'_H)=-1$ and $\cos(\chi''+2\delta'_V)=-1$, and then define
\begin{align}\label{P-MT-max-supp}
P_\mu^{\prime (-)}=\left(P_\mu\big|_{\theta=0}^{\delta=\delta'_\mu}\right)_{\text{max}} -\left(P_\mu\big|_{\theta=0}^{\delta=\delta'_\mu}\right)_{\text{min}},
\end{align}
where $\mu=H,V$. 
\par
Using Eqs.~(\ref{ph-countingrate-h-supp}), (\ref{ph-countingrate-v-supp}), (\ref{P-MT-supp:a}), (\ref{P-MT-supp:b}), and (\ref{P-MT-max-supp}), we obtain
\begin{subequations}\label{P-M-HV-supp}
	\begin{align}
	P_H^{(+)}&=\frac{1-\eta}{2}+\eta I_H, \label{P-M-HV-supp:a} \\
	P_H^{(-)}&=2|b_1||b_2|\left|(T_H-T_V)\frac{1-\eta}{4}+T_H \eta I_H\right|, \label{P-M-HV-supp:b} \\
	P_V^{(+)}&=\frac{1-\eta}{2}+\eta I_V, \label{P-M-HV-supp:c} \\
	P_V^{(-)}&=2|b_1||b_2|\left|(T_V-T_H)\frac{1-\eta}{4}+T_V \eta I_V\right| \label{P-M-HV-supp:d}, \\
	P_H^{\prime (-)}&=2|b_1||b_2|\left((T_H+T_V)\frac{1-\eta}{4}+T_H \eta I_H\right), \label{P-M-HV-supp:e} \\
	P_V^{\prime (-)}&=2|b_1||b_2|\left((T_V+T_H)\frac{1-\eta}{4}+T_V \eta I_V\right)\label{P-M-HV-supp:f},
	\end{align}
\end{subequations}
where we recall that $I_V=1-I_H$. 
\par
From the set of equations (\ref{P-M-HV-supp:a})--(\ref{P-M-HV-supp:f}), we can always find four linearly independent equations [e.g., (\ref{P-M-HV-supp:b}), (\ref{P-M-HV-supp:d}), (\ref{P-M-HV-supp:e}), and (\ref{P-M-HV-supp:f})] that involve four unknowns, $T_H$, $T_V$, $\eta$, and $I_H$. Therefore, it is always possible to obtain unique solutions for $T_H$ and $T_V$, i.e., experimental imperfections can be quantitatively characterized from the measurement data. We, however, do not go into the details of finding $T_H$ and $T_V$ since it is an experimental problem. The fact that $T_H$ and $T_V$ can be determined is enough for our purpose. We treat $T_H$ and $T_V$ as experimentally measurable quantities.
\par
If $T_H\geq T_V$, we find using Eqs.~(\ref{b1b2-supp}), (\ref{P-M-HV-supp:a}), and (\ref{P-M-HV-supp:b}) that
\begin{align}\label{eta-supp-1}
\eta=\frac{2P_H^{(-)}+|b_1||b_2| (T_{H}+T_{V}-4P_H^{(+)}T_{H})}{|b_1||b_2|(T_{V}+T_{H})}.
\end{align}
Similarly, if $T_H\leq T_V$, we obtain the following expression using Eqs.~(\ref{b1b2-supp}), (\ref{P-M-HV-supp:c}), and (\ref{P-M-HV-supp:d}) as follows,
\begin{align}\label{eta-supp-2}
\eta=\frac{2P_V^{(-)}+|b_1||b_2|(T_{V}+T_H-4P_V^{(+)}T_{V})}{|b_1||b_2|(T_{H}+T_{V})}.
\end{align}
It can be checked using Eqs.~(\ref{P-M-HV-supp:a})--(\ref{P-M-HV-supp:d}) that when $T_H=T_V$, Eqs.~(\ref{eta-supp-1}) and (\ref{eta-supp-2}) become identical to each other. Combining Eqs.~(\ref{eta-supp-1}) and (\ref{eta-supp-2}), we express $\eta$ in terms of experimentally measurable quantities as
\begin{equation}\label{eta-supp}
\eta =
\begin{cases}
\frac{2P_H^{(-)}+|b_1||b_2| (T_{H}+T_{V}-4P_H^{(+)}T_{H})}{|b_1||b_2|(T_{V}+T_{H})}& \text{if $T_H \geq T_V$},\\[5 pt]
\frac{2P_V^{(-)}+|b_1||b_2| (T_{V}+T_H-4P_V^{(+)}T_{V})}{|b_1||b_2|(T_{H}+T_{V})} & \text{if $T_H \leq T_V$},
\end{cases}       
\end{equation}
where an expression for $|b_1||b_2|$ in terms of experimentally measurable quantities is given by Eq.~(\ref{b1b2-supp}).
\begin{table*}[t]		
	\setlength{\tabcolsep}{10pt} 
	\renewcommand{\arraystretch}{1} 
	\begin{tabular}{c  c  c  c  c  c} 
		\hline\hline 
		State & ($\eta,\I,I_H$) &
		$\widetilde{P}_{HV}$ &
		$\widebar{P}_{DR}$ &
		PPT Criterion & Concurrence \\[2 pt] 
		\hline 
		$\widehat{\rho}_1$ & (0.0, --, --) & 0.00 & 0.00 & Separable & 0.00 \\
		$\widehat{\rho}_2$ & (0.2, 1.0, 0.5) & 0.20 & 0.10 & Separable & 0.00 \\
		$\widehat{\rho}_3$ & (0.6, 0.8, 0.3) & 0.6 & 0.22 & Entangled& 0.24 \\ 
		$\widehat{\rho}_4$ & (0.7, 1.0, 0.5) & 0.69 & 0.35 & Entangled & 0.55  \\
		$\widehat{\rho}_5$ & (1.0, 1.0, 0.5) & 1.00 & 0.50 & Entangled & 1.00 \\ 
		\hline\hline
	\end{tabular}
	\caption{Numerical results illustrating entanglement measurement in the presence of experimental imperfections. Imperfections are characterized by choosing $T_H=0.25$ and $T_V=0.35$.}\label{tab:1-supp}
\end{table*}
\par
We now set $\theta=\pi/4$ in Eqs.~(\ref{ph-coutng-rate-DARL-supp:a})--(\ref{ph-coutng-rate-DARL-supp:d}). In analogy with Eq.~(\ref{P-MT-supp:b}), we define
\begin{align}\label{P-M-DARL-supp}
P_\nu^{(-)}&=\left(P_\nu\big|_{\theta=\frac{\pi}{4}}\right)_{\text{max}}-\left(P_\nu\big|_{\theta=\frac{\pi}{4}}\right)_{\text{min}},
\end{align}
where $\nu=D,A,R,L$. Now from Eqs.~(\ref{ph-coutng-rate-DARL-supp:a})--(\ref{ph-coutng-rate-DARL-supp:d}) and (\ref{P-M-DARL-supp}), we have
\begin{subequations}\label{P-M-DARL}
	\begin{align}
	&P_D^{(-)}=P_A^{(-)}=\sqrt{T_H^2+T_V^2+2T_HT_V\cos (\chi'-2\delta)} \nonumber \\ & \qquad \qquad \qquad \times |b_1||b_2| \eta \mathscr{I}\sqrt{I_HI_V}\label{P-M-DARL:a},\\
	&P_R^{(-)}=P_L^{(-)}=\sqrt{T_H^2+T_V^2-2T_HT_V\cos (\chi'-2\delta)} \nonumber \\ & \qquad \qquad \qquad \times |b_1||b_2| \eta \mathscr{I}\sqrt{I_HI_V}\label{P-M-DARL:b}.
	\end{align}
\end{subequations}
Solving Eqs.~(\ref{P-M-DARL:a}) and (\ref{P-M-DARL:b}), we get
\begin{align}\label{P-DR-supp-main}
\eta\I\sqrt{I_HI_V}=\frac{\sqrt{2\big[(P_D^{(-)})^2+(P_R^{(-)})^2}\big]}{2|b_1||b_2|\sqrt{T_{H}^2+T_{V}^2}},
\end{align}
where $|b_1||b_2|$ is given by Eq.~(\ref{b1b2-supp}).
\par
To test the PPT criterion, we express the eigenvalue $\alpha_1$ [Eq.~(2) in main paper] in terms of experimentally measurable quantities. Using Eq.~(2) from the main paper and Eqs.~(\ref{b1b2-supp}), (\ref{eta-supp}), and (\ref{P-DR-supp-main}), we find that 
\begin{align}
\alpha_1=\frac{1-\widetilde{P}_{HV}(T_H,T_V)-4\widebar{P}_{DR}}{4},
\end{align}
where $\widetilde{P}_{HV}$ and $\widebar{P}_{DR}$ are right-hand sides of Eqs.~(\ref{eta-supp}) and (\ref{P-DR-supp-main}) respectively.
According to the PPT criterion, the state $\dm$ [Eq.~(\ref{q-state-supp})] is entangled if and only if $\alpha_1<0$, that is, the state is entangled if and only if
\begin{align}\label{PPT-prob-supp}
\widetilde{P}_{HV}(T_H,T_V)+4\widebar{P}_{DR}>1.
\end{align}
It can be readily checked that Eq.~(\ref{PPT-prob-supp}) reduces to Eq.~(E6) of the main paper when $T_H=T_V=1$. That is it becomes equivalent to  Eq.~(21) of the main text when there is no experimental imperfections. 
\par
To numerically illustrate our results, we choose the same five quantum states ($\dm_1,\dots,\dm_5$) considered in Table I of the main paper. We consider a high loss scenario in which $T_H=0.25$ and $T_V=0.35$. We test whether these states are entangled using Eq.~(\ref{PPT-prob-supp}) and present the results in Table \ref{tab:1-supp}. We find that the results are identical to those found assuming the absence of experimental imperfections.
\par
In order to express the concurrence in terms of experimentally measurable quantities, we substitute from Eqs.~(\ref{b1b2-supp}), (\ref{P-DR-supp-main}) and (\ref{eta-supp}) into Eq.~(3) of the main paper and find that
\begin{align}\label{conc-prob-supp}
C(\widehat{\rho})=\text{max}\Big\{0,\frac{\widetilde{P}_{HV}(T_H,T_V)+4\widebar{P}_{DR}-1}{2}\Big\},
\end{align}
where $\widetilde{P}_{HV}$ and $\widebar{P}_{DR}$ are right-hand sides of Eqs.~(\ref{eta-supp}) and (\ref{P-DR-supp-main}) respectively. It can be readily checked that when $T_H=T_V=1$, Eq.~(\ref{conc-prob-supp}) reduces to Eq.~(E7) of Appendix E of the main paper. That is, it is fully equivalent to Eq.~(22) of the main paper in absence of experimental imperfections.
\par
We numerically illustrate the formula for the concurrence using the same five quantum states used for testing the PPT criterion. Using Eq.~(\ref{conc-prob-supp}), we determine the concurrence of these states for $T_H=0.25$ and $T_V=0.35$. The results are displayed in the rightmost column of Table \ref{tab:1-supp}. They are identical to those obtained assuming no experimental imperfections in the main paper.
\par
Our results thus establish that the technique is resistant to experimental imperfections. 
	
\end{document}